\documentclass[10pt,journal,compsoc]{IEEEtran}\usepackage[]{graphicx}\usepackage[dvipsnames]{xcolor}
\makeatletter
\def\maxwidth{ %
  \ifdim\Gin@nat@width>\linewidth
    \linewidth
  \else
    \Gin@nat@width
  \fi
}
\makeatother

\definecolor{fgcolor}{rgb}{0.345, 0.345, 0.345}

\usepackage{framed}
\makeatletter
 {\par\unskip\endMakeFramed%
 \at@end@of@kframe}
\makeatother

\definecolor{shadecolor}{rgb}{.97, .97, .97}
\definecolor{messagecolor}{rgb}{0, 0, 0}
\definecolor{warningcolor}{rgb}{1, 0, 1}
\definecolor{errorcolor}{rgb}{1, 0, 0}
\newenvironment{knitrout}{}{} % an empty environment to be redefined in TeX

\usepackage{alltt}

\usepackage[T1]{fontenc}
\usepackage{helvet}
\usepackage{cite}
\usepackage{url}
\usepackage{hyperref}
\usepackage{graphicx}
\usepackage{enumitem}  % For letter-based lists
\usepackage{listings}  % To show code/text as an example
\usepackage{booktabs}

\graphicspath{ {../img} {img} {.} {tex} }  % workdir can be 'tex' or 'tex/..'
\usepackage[dvipsnames]{xcolor}
\usepackage{balance}
\newcommand{\ifwithappendix}[1]{}  % activate this for the journal version of the document (no appendix)

\newcommand{\PlotCustom}[5]{%
	\begin{#1}[#3]%
		\centering\includegraphics[width=#4]{#2}%
		\vspace{-4mm}\caption{#5}\label{#2}%
	\end{#1}}
\newcommand{\Plot}[2]{\PlotCustom{figure}{#1}{htb}{\columnwidth}{#2}}
\newcommand{\PlotWidth}[3]{\PlotCustom{figure}{#1}{htb}{#2\columnwidth}{#3}}
\newcommand{\Plotwide}[2]{\PlotCustom{figure*}{#1}{tbp}{\textwidth}{#2}}

\newcommand{\Cb}[1]{\bgroup\mbox{\scshape #1}\egroup}  % codebook entry name
\newcommand{\Prg}[1]{\bgroup\ttfamily #1\egroup}  % commands and script names
\newcommand{\strct}[1]{\bgroup\ttfamily\bfseries #1\egroup}  % abstract structure topicletter string
\newcommand{\Art}[1]{\href{https://github.com/serqco/qabstracts/blob/main/abstracts/abstracts.A/\string#1.txt}{\bgroup[#1]\egroup}}
\newcommand{\Repo}[2]{\href{https://github.com/serqco/qabstracts/blob/main/\string#2}{#1}}
\newcommand{\Sah}[1]{\bgroup\scshape #1\egroup}  % structured abstract heading

\newcommand{\codebook}{https://github.com/serqco/qabstracts/blob/main/codebook.md}

\definecolor{todocolor}{rgb}{1, 0, 0}
\definecolor{abstractcolor}{RGB}{191, 27, 73}
\definecolor{codebkecolor}{RGB}{0, 16, 218}
\definecolor{pseudocolor}{RGB}{0, 144, 83}

\newcounter{todonumber}

\newcommand{\Quote}[1]{\bgroup\itshape\color{abstractcolor} ``#1''\egroup}   % quote from a specific real abstract
\newcommand{\ExampleQuote}[1]{\bgroup\itshape ``#1''\egroup}                 % generic quote from real abstract
\newcommand{\Pseudoquote}[1]{\bgroup\itshape\color{pseudocolor} `#1'\egroup} % quote from fictive abstract
\newcommand{\QuoteCB}[1]{\bgroup\itshape\color{codebkecolor} ``#1''\egroup}  % quote from codebook
\newcommand{\Describeboxplots}{The box shows the 25-to-75 percentile, 
	the whiskers are 10- and 90-percentile,
	the gray bar is the median,
	the fat dot is the mean.}
\newcommand{\Describegroups}{The plots in each group show these different subsets of abstracts from left to right:
	all (red); structured, non-structured (green); design, empirical (blue); EMSE, ICSE, IST, TOSEM, TSE (gray).}
\IfFileExists{upquote.sty}{\usepackage{upquote}}{}
\begin{document}

\title{How (Not) To Write a\\Software Engineering Abstract}

\author{Lutz Prechelt%
\thanks{L. Prechelt is with Freie Universität Berlin, Germany},
Lloyd Montgomery%
\thanks{L. Montgomery is with University of Hamburg, Germany},  % Our university prefers the English name in publications
Julian Frattini%
\thanks{J. Frattini is with Chalmers University of Technology and University of Gothenburg, Sweden},
Franz Zieris%
\thanks{F. Zieris is with Blekinge Institute of Technology (BTH), Karlskrona, Sweden}}

\markboth{IEEE Transactions on Software Engineering}%
{How (Not) To Write a Software Engineering Abstract}

\maketitle

%%%%%%%%%%%%%%%%%%%%%%%%%%%%%%%%%%%%%%%%%%%%%%%%%%%%%%%%%%%%%%%%%%%%%%%%%%%%%%%%
%                                ARTICLE
%%%%%%%%%%%%%%%%%%%%%%%%%%%%%%%%%%%%%%%%%%%%%%%%%%%%%%%%%%%%%%%%%%%%%%%%%%%%%%%%

% Note: Make sure to update `our-abstract.txt', from which our own FK score gets calculated in 
%       the caption of Fig. 3.
%       The headers "Background:" etc. must be on a line of their own with no additional whitespace.
\begin{abstract}  % 100 to 200 words are allowed (but not enforced)
\emph{Background:}
Abstracts are a particularly valuable element in a software engineering research article.
However, not all abstracts are as informative as they could be.
\emph{Objective:}
Characterize the structure of abstracts in high-quality software engineering venues.
Observe and quantify deficiencies.
Suggest guidelines for writing informative abstracts.
\emph{Methods:}
Use qualitative open coding to derive concepts that explain relevant properties of abstracts.
Identify the archetypical structure of abstracts.
Use quantitative content analysis to objectively characterize abstract structure
of a sample of 362 abstracts from five presumably high-quality venues.
Use exploratory data analysis to find recurring issues in abstracts.
Compare the archetypical structure to actual structures.
Infer guidelines for producing informative abstracts.
\emph{Results:}
Only 29\% of the sampled abstracts are \textit{complete},
i.e., provide background, objective, method, result, and conclusion information.
For structured abstracts, the ratio is twice as big.
Only 4\% of the abstracts are \textit{proper},
i.e., they also have good readability (Flesch-Kincaid score) and have no 
informativeness gaps, understandability gaps, nor highly ambiguous sentences.
\emph{Conclusions:}
(1)~Even in top venues, a large majority of abstracts are far from ideal.
(2)~Structured abstracts tend to be better than unstructured ones.
(3)~Artifact-centric works need a different structured format.
(4)~The community should start requiring conclusions that generalize,
which currently are often missing in abstracts.
\end{abstract}


%========================================================================
\section{Introduction}
\IEEEPARstart{A}{lthough} the abstract is a super important part of any 
research article~\cite{Lang22,ShiGalMil24},
when reading an abstract in software engineering --- even in a presumably top-quality venue ---
we often feel it is lacking important information or find it difficult to understand at all.

We aim to substantiate this impression
by operationalizing \emph{abstracts quality} and analyzing hundreds of abstracts.
We want to formulate constructive advice for writing better abstracts.

\subsection{Research Questions}

% we use the numbers only in this subsection and the next;
% everywhere else we repeat the question or paraphrase it.
\noindent
We ask several questions:
\begin{description}
	\item[RQ1] What does a typical well-written abstract look like?
	\item[RQ2] \textbf{a)} Which deficiencies occur? \textbf{b)} How often?
	\item[RQ3] Do structured abstracts have better quality than unstructured ones?
	\item[RQ4] How \emph{should} software engineering abstracts be written?
\end{description}
We consider RQ1 and RQ2 to be exploratory.
RQ3 is hypothesis-driven; we expect a `yes'.
The answer to RQ4 will be derived from the answers to the other three.

\subsection{Research Approach}

No firm expectations regarding RQ1 and RQ2a exist for engineering articles,
so qualitative methods will have to be used for them:
We start our research with \emph{open coding} of software engineering abstracts
to derive a vocabulary (a set of concepts or \emph{codes}; \cite[Ch. 5]{StrCor90})
by which the nature of a particular abstract can be characterized.

We intend to convince even readers who are skeptical of qualitative research
regarding our answers to the research questions and the need to improve
the quality of software engineering abstracts.
We therefore perform a repeatable
quantitative content analysis~\cite[Ch. 7]{Krippendorff04}
on a large sample of 362 presumably high-quality abstracts.
We apply an elaborate eight-step approach for maximizing reliability.

The final stage is a statistical evaluation of the content analysis data,
which is again exploratory and straightforward in
the questions it asks and the statistical methods it applies.

\subsection{Research Contributions}

Our contributions correspond to the research questions as follows:
\begin{description}
	\item[RQ1] As for the structure of well-written abstracts, we present an ``abstracts archetype''
	  that describes fixed parts of the structure and degrees of freedom
	  (Section~\ref{archetype}).
	\item[RQ2] We describe and discuss eight types of deficiencies;
	  we quantify the frequency of each deficiency type for the entire sample of abstracts
	  as well as for different subgroups of interest
    (Section~\ref{results} and Table~\ref{tab:overview}).
	  % Deficiency Types:
	  %  1. Readability (Sec 4.1, result-len-read)
	  %  2. Inefficient Allocation of Space (Sec 4.4, inefficient_allocation)
	  %  3. Missing Elements (Sec 4.6, missing_elements)
    %  4. Confusing abstracts (Sec. 4.7, confusing)
    %  5. Informative Gaps (Sec. 4.8.1, igaps)
    %  6. Announcements (Sec. 4.8.2, announcements)
    %  7. Understandability (Sec. 4.9, ugaps)
    %  8. Ambiguity (Sec. 4.10, ambiguity)
	\item[RQ3] We present convincing data that structured abstracts tend to be better
	  in several respects
	  (Sections~\ref{missing_elements} \& \ref{structuredgood}).
	\item[RQ4] We provide data-based how-to instructions for writing abstracts for authors;
	  we provide guidance for editors and conference organizers
    (Section~\ref{howto-guidelines}).
	  Software engineering works should use structured abstracts, but
	  need a different and more flexible template than what is used so far.
\end{description}

%========================================================================
\section{Related Work}

There is considerable literature on research abstracts across disciplines.
We will not attempt to summarize it here,
but provide examples of the different major perspectives of those studies.
We otherwise focus on what has been done in the software engineering domain.

\subsection{Abstracts Structure}

Swales~\cite{Swales90} introduced \emph{genre analysis} as a means for teaching academic
reading and writing, especially to non-native speakers:
Genres are ``classes of communicative events'' (e.g., \emph{the writing and reading of abstracts})
that are owned by a ``discourse community'' (e.g., \emph{software engineering researchers});
genre analysis means deconstructing texts (from a genre) to better understand
their elements in
terms of their syntactical structure, content, role, and interrelationships
(e.g., their relative position within the whole).

For our purposes here, the most relevant idea from genre analysis is the notion
of ``moves'', which are, roughly speaking, the building blocks used by writers
for making their overall point.
Several studies have looked at the move structure of research abstracts
in different fields such as
applied linguistics~\cite{DosSantos96} or
protozoology~\cite{CroOpp06}.
Despite the differences of research content, they find very similar moves,
typically the following five-move structure~\cite{CroOpp06}:
\begin{quote}
	(1)~situate the research within the scientific community;
	(2)~introduce the research by describing the main features or presenting its purpose;
	(3)~describe the methodology;
	(4)~state the results;
	(5)~draw conclusions or suggest practical applications.
\end{quote}
This reflects, in slightly extended form, the IMRAD structure 
(Introduction, Methods, Results, and Discussion) of the body of scientific articles
that has gradually become the norm since the 1940s \cite{SolPer04}.

We found a similar structure for abstracts of 
empirical works in software engineering (see Section~\ref{archetype}),
but abstracts of artifact-centric works (tool building) do not fit this model and need
an extended one \ref{structure-design-articles}.

\subsection{Abstracts Quality -- What is ``good''?}\label{relatedworkquality}

Several (meta-)studies on abstracts focus on quality assessment,
most often in subfields of the biomedical domain.
Many such studies cover articles of a homogeneous nature:
randomized controlled trials (controlled experiments).
This allows formulating specific expectations regarding what information should be
presented in an abstract and allows performing the analysis in checklist fashion,
for example:
In clinical dermatology, \cite{DupKhoLeb03} used a 30-item checklist on 197 abstracts
  for computing a 0-to-1 completeness score and found mean scores between 
  0.64 and 0.78 for their various subgroups.
In dental medicine, \cite{ShaHar06} used a 29-item checklist on 100 abstracts
  and found a mean score of only 0.54.
  % suggests an 8-move structure for structured abstracts:
  % objective, design, setting, patients, interventions, outcome measure, results, conclusion.
  % per-move deficit rates are not reported.
Among 303 abstracts of cost-effectiveness analyses,
  29\% did not report the baseline to which the intervention
  had been compared~\cite{RosGreSto05}.
Among 146 abstracts of meta-analyses in peridontology, 
  33\% did not even report the direction in which  
  % it is indeed 32.9% percent: (146-15-83)/146, Table 2
  the evidence was pointing~\cite{FagLiuHud14}.
  
Various studies have investigated ``spin'' in the context of significance testing.
The term covers two types of behavior: 
Using language that sounds more positive than warranted or
reporting a secondary or alternative statistic as if it was the main one of interest.
Among all abstracts reporting non-significant results, studies found spin in 
  45\% of abstracts of orthopedic controlled experiments~\cite{ArtZaaChe20},
  44\% in emergency medicine~\cite{ReyRidBro20},
  56\% in psychiatry and psychology~\cite{JelRobBow20},
  and 58\% for the conclusions alone across a broad set of 
    medical controlled experiments~\cite{BouDutRav10}.

Unfortunately, the methods of those meta-studies are not applicable to a 
broad sample of software engineering abstracts, because
in fields with heterogeneous study structures such as ours, the operationalization of
\emph{quality} is less straightforward.
For example, \emph{spin} can take many more forms in software engineering articles
than is assumed by the studies mentioned above. It is difficult
to decide which forms are acceptable and which are not.

One approach for discussing the quality of a software engineering abstract 
could be through comparing to a known ``good'' structure for abstracts.
For instance,~\cite{CroOpp06} remarks
that one third of the 12 analyzed abstracts is lacking move 2 (stating a purpose).

\subsection{Structured vs.\ Unstructured Abstracts}\label{relatedworkstructured}

Many studies of abstracts quality do not study quality in general.
For instance, neither~\cite{DupKhoLeb03} nor~\cite{ShaHar06} reports
which of their checklist items are missing most frequently.
Rather, their research question is the relative quality of structured
versus unstructured abstracts.
Definition: A \emph{structured abstract} is one that uses a prescribed sequence of 
intermediate headings, such as Background, Objective, Methods, Results,
Conclusions or some similar set.
In almost all of those meta-studies, including~\cite{DupKhoLeb03} and~\cite{ShaHar06},
the answer is: structured abstracts have fewer quality issues.
Such research can be highly influential.
For instance, the CONSORTS report~\cite{MohHopSch12}
(containing guidelines for reporting controlled experiments, with over 10,000 citations),
relies on such a study \cite{HarSydBlu96} to recommend structured abstracts.
% https://journals.sagepub.com/doi/epdf/10.1177/016555159602200503

In software engineering, Kitchenham proposed
\emph{Evidence Based Software Engineering} (EBSE) in 2004 \cite{KitDybJor04}.
EBSE relies a lot on Systematic Literature Reviews (SLR).
The practicality of SLRs hinges on the informativeness of abstracts:
Can the researcher decide quickly and reliably, whether the present article
belongs in the SLR or not?

Therefore, Kitchenham performed two studies on structured abstracts
in software engineering.
The first took 23 published non-structured abstracts, 
converted them into structured ones,
and compared the two versions. 
It found that the structured abstracts were much longer,
but also had much better readability scores \cite{KitBreOwe08}.
The second, by Budgen, Kitchenham, and others \cite{BudKitCha08},
is a controlled experiment based on similar pairs of abstracts
rewritten into the structure 
Background, Aim, Methods, Results, Conclusions.
20 students and 44 researchers and practitioners
each judge one structured and one different unstructured abstract
for completeness (using an 18-item checklist) and 
clarity (using a vague 1-to-10 scale).
The structured format was found to increase the 
completeness score by 6.6 and the clarity score by 3.0.
70\% of the subjects also preferred the structured format
subjectively.
Both studies use only abstracts of purely empirical studies,
not tool-building works, for which structured abstracts require and extended form
as we will see in Section~\ref{structure-design-articles}.

%========================================================================
\section{Methods}

\subsection{Overview}

Our study is a full-blown content analysis in the sense of Krippendorff~\cite{Krippendorff04}:
not just a counting exercise with a fixed codebook,
but rather an iterative codebook development before (and during) the counting
and an extensive abductive inference exercise after the counting.
It can be conceptualized as consisting of four widely overlapping stages or phases
as shown in Figure~\ref{qabstracts_timeline_commits}:
\Plot{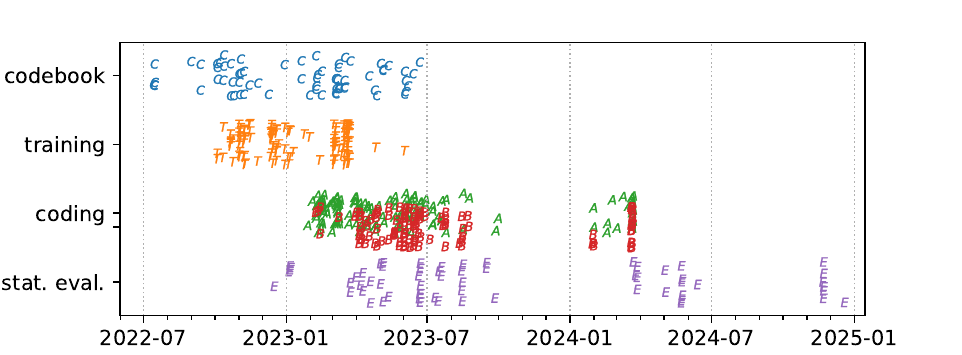}{%
	Timeline of the main study phases and their individual events.
    Each character's x-coordinate represents the time of a git commit.
    The vertical scattering is added for legibility only.}
Codebook development, which started first and is described in Sections~\ref{meth_codingrules}
and \ref{meth_codebook},
defined the rules and target concepts of the counting.
Training (Section~\ref{meth_training}) was for developing a joint understanding of the codebook.
It also contributed greatly to the codebook's early evolution.
Coding worked on a large sample of abstracts from top-quality venues described in
Section~\ref{meth_sample} and produced the count data subsequently used in the statistical analysis.
Our coding process and our subsequent statistical evaluation are
described in Sections~\ref{meth_coding} and \ref{statisticalevaluation}.

Overall, we consider our study to be a \emph{qualitative} one,
but coding and statistical evaluation are also largely
compatible with a \emph{positivist} epistemology (aiming for ``objective'' results),
such that the final interpretation is done on a solid \emph{quantitative} foundation.

\subsection{Data Availability}\label{dataavailability}

We publish not only the outcome of our study, but also most parts of its development and
execution history in full detail: as a git version repository.
It includes all versions of the
codebook, handling procedure, coded abstracts, Python scripts for automation,
Python scripts for tabulations and plots, and the manuscript of this article in 
Knitr LaTeX format (that is, with automated computation of most of the numbers in the text).
Find it at \url{https://github.com/serqco/qabstracts/}.
A snapshot of the most important files is available at Zenodo as
\cite{PreMonFra25}.

When we refer to abstracts from our sample, we use abbreviations of the first three authors' names and the year of publication,
e.g., \Art{BesMarBos22}. Such strings are hyperlinks to the annotated abstract in our GitHub repository.
Refer to the repository in order to find the corresponding bibliographic information.

\subsection{General Coding Rules}\label{meth_codingrules}

Besides the definition of the content categories (codes),
our codebook contains global rules for the coding that can be summarized as follows:
\begin{itemize}
\item In order to limit complexity and avoid arbitrariness, we code by sentence and
  prefer single codes per sentence over multi-codings.
\item In order to mimic ordinary readers, when choosing a code, 
  we consider only what we have seen before plus
  one sentence forward context when needed (and only when needed).
\item In order to avoid excessive criticism of abstracts quality,
  we avoid coding negative properties whenever 
  an alterative, more positive interpretation is plausible as well.
\end{itemize}

\subsection{Codebook and Codebook Development}\label{meth_codebook}

\subsubsection{\Cb{background}, \Cb{objective}, \Cb{method}, \Cb{result}, \Cb{conclusion}}\label{meth_basicparts}

The \Repo{codebook}{codebook.md} 
was initialized with concepts for the five sections commonly used
in a structured abstract:\footnote{In Krippendorff's terminology, this is an ``established theories'' justification
of analytical constructs~\cite[Section 9.2.3]{Krippendorff04}.}
\Cb{background, objective, method, result, conclusion}.
These represent, respectively: context information, the study goal or question,
empirical approach, empirical outcomes, and a take-home message that generalizes beyond the results.

Each code in the codebook is defined by a short verbal explanation.
For example, we defined \Cb{method} as
\QuoteCB{information about the approach or setup of an empirical (or possibly purely mathematical) study}.
The initial definitions were made more and more precise and unambiguous by later
codebook refinements.
The part \QuoteCB{(or possibly purely mathematical)} is such a clarification
that we added when we encountered the first of those (very few) purely mathematical studies
during the coding process and were confused which code was appropriate.
To help disambiguation, a few of the definitions need to be much longer than the above.

\subsubsection{Artifact-centric Studies: \Cb{design}}\label{design}

We then had a phase (internally called ``prestudy'') when we refined and extended the codebook
based on coding attempts for a stratified sample of 20 abstracts from ICSE 2021.
% we used every 7th article, which constitutes stratification by virtue of the
% topic-centric session blocks in which the program is arranged.
We quickly recognized that additional codes were
needed.\footnote{In Krippendorff's terminology, this is an ``expert knowledge and experience'' justification
    of analytical constructs~\cite[Section 9.2.2]{Krippendorff04}:
    Being software engineering researchers ourselves, 
    we recognize when a sentence makes a different kind
    of contribution to an abstract than can be described by existing codes,
    and we are able to define what kind of contribution it is.}
Most importantly,
many software engineering articles do not talk about only an empirical study.
Rather, their focus is the design of some artifact, most often a tool,
sometimes a method or something else.
Much of the abstract is then spent on design considerations, design decisions,
techniques applied in implementation, and so on.
Such artifact-centric studies, although they also usually also contain an empirical study,
are quite different in nature from purely empirical studies;
we therefore introduced the code \Cb{design} to mark such material.\footnote{\Cb{design} has
  the most detailed definition of all our codes: 170 words.}

We call the articles that contain at least one \Cb{design} coding in their abstract
``design articles'', the others ``empirical articles''.

\subsubsection{Refinements: \Cb{gap}, \Cb{summary}, \Cb{fposs}, etc.}

At various later points during our study, we recognized a need for more granular coding
in order to capture differences in abstract writing we wanted to measure.
This led to the splitting of existing codes and the introduction of additional ones.
We then reworked existing codings to use the new codes consistently throughout.

The most important cases of such additional codes are these:
\Cb{gap} states what is unknown or not yet possible;
\Cb{summary} summarizes several results, but does not provide new information
(whereas \Cb{conclusion} generalizes beyond the immediate results);
\Cb{fposs} and \Cb{fneed} sentences occur at the end of abstracts ---
they state what future work is now possible or needed.

\subsubsection{Codes for Announcements: \Cb{a-*}}
\label{announcement-codes}

Sometimes, a statement in an abstract insinuates there will be certain information 
in the article body, but does not provide any concrete information itself.
We call such statements ``announcements'' and use codes with an \Cb{a-*} prefix for them.
For example, here is an \Cb{a-method} (method announcement) from \Art{BesMarBos22}:
\Quote{As a second step, this study sets out to specifically
  provide a detailed assessment of additional and in-depth analysis of technical debt management strategies based
  on an encouraging mindset and attitude from both managers and technical roles to understand how, when and by
  whom such strategies are adopted in practice.}
Despite the long sentence, we learn nothing about how the assessment
or the analysis work.
Here is an \Cb{a-result} (results announcement) from \Art{FlyChaDye22}:
\Quote{Based on those results, we then used the Boa and Software
  Heritage infrastructures to help identify and quantify several sources of dirty Git timestamp data.}
The first part is \Cb{method}, the second should have been \Cb{result},
but, alas, we learn nothing about those sources' nature or number or impact.

\subsubsection{Codes for Headings: \Cb{h-*}}

We use \Cb{h-*} codes for the headings used in structured abstracts.
These codes are conceptual, i.e., for instance
\ExampleQuote{Aim:}, \ExampleQuote{Goal:}, \ExampleQuote{Objective:}, \ExampleQuote{Question:},
and their plural forms
would all be coded as \Cb{h-objective}.
Likewise, there are \Cb{h-background}, \Cb{h-method}, and so on.
We consider an abstract to be structured if it has at least one such heading code.

\subsubsection{Subjective Additions: \Cb{:i, :u}}
\label{subjective-codes}

The codes described so far aim at codifying repeatable properties,
where several well-trained coders will come to the same result with high probability.
In addition, we defined a number of suffixes for codes, by which coders can provide
additional information for which the expectation of agreement is much lower.
These are also not neutral, like the codes themselves, but all describe some
kind of deficiency.
The most important of these suffixes are the following:

Informativeness gaps (coded as \Cb{:i})
are spots in a sentence where the coder desired to know
additional detail that is presumably available to the authors and
that can presumably be provided in very little space.
Example (from \Art{LiuFenYin22}):
\Quote{To evaluate DeepState, we conduct an extensive empirical study on popular datasets
and prevalent RNN models containing image and text processing tasks.}
This sentence was coded as \Cb{method:i2}, because the coder asked themselves
\Pseudoquote{How many datasets? How many RNN models?}
The answers to both are given in that article's Table 3: four datasets, three models.
The authors could and should have given that information in the abstract.

Understandability gaps (coded as \Cb{:u})
are spots in a sentence where the coder encountered a term they did not know
and found that the intuitive partial understanding they could muster for that term
to be insufficient for understanding the abstract overall.
Example (from \Art{CheHuWei22}):
\Quote{Finally, we propose a new dynamic vocabulary strategy which can effectively resolve the 
  UNK problems in code summaries.}
This sentence was coded as \Cb{design:u1}
because it is unclear from the abstract what these apparently important \Quote{UNK problems} are
supposed to be.
% Here are all of the other candidate "easy-to-understand examples of :u1" that I found:
% The data were analyzed using a Bayesian independent sample t-test and **network
% analysis**.
% {{method:u1}}
% The evaluation results suggest that their edge coverage performance can be **unstable**.
% {{result:u1}}
% Our results motivate short-term patches and **long-term fundamental solutions**.
% {{conclusion:u1}}
% In this paper, we propose **CDCS**, a novel approach for domain-specific code search.
% {{objective:u1}}

\begin{knitrout}\scriptsize
\definecolor{shadecolor}{rgb}{0.969, 0.969, 0.969}\color{fgcolor}\begin{table*}

\caption{\label{tab:codes}Topics, codes, their descriptions
      (see our \href{\codebook}{codebook} for full descriptions),
      and how often we used them to code sentences.}
\centering
\begin{tabular}[t]{llr}
\toprule
\textbf{\emph{Topic} / Code} & \textbf{Short Description} & \textbf{Occurrences}\\
\midrule
\addlinespace[0.3em]
\multicolumn{3}{l}{\textbf{\emph{Background} (\texttt{b})}}\\
\hspace{1em}\Cb{background} & Context information & 2003\\
\hspace{1em}\Cb{h-background} & Heading (e.g., \ExampleQuote{Background:} or \ExampleQuote{Context:}) & 164\\
\addlinespace[0.3em]
\multicolumn{3}{l}{\textbf{\emph{Gap} (\texttt{g})}}\\
\hspace{1em}\Cb{gap} & Unknown or not-yet-possible things, leading over to \Cb{objective} & 538\\
\hspace{1em}\Cb{need} & Research that needs to be done (postulated), leading over to \Cb{objective} & 26\\
\addlinespace[0.3em]
\multicolumn{3}{l}{\textbf{\emph{Objective} (\texttt{o})}}\\
\hspace{1em}\Cb{objective} & Top-level research goal, interest, or question & 907\\
\hspace{1em}\Cb{h-objective} & Heading (e.g., \ExampleQuote{Objective:} or \ExampleQuote{Aim:}) & 162\\
\addlinespace[0.3em]
\multicolumn{3}{l}{\textbf{\emph{Design} (\texttt{d})}}\\
\hspace{1em}\Cb{design} & Design, design process or features of an artifact (e.g., software, process, method) & 1091\\
\hspace{1em}\Cb{a-design} & \Cb{design} announcement (e.g., \ExampleQuote{A description of the tool is presented.}) & 18\\
\addlinespace[0.3em]
\multicolumn{3}{l}{\textbf{\emph{Method} (\texttt{m})}}\\
\hspace{1em}\Cb{method} & Approach or setup of an empirical (or possibly purely mathematical) study & 1172\\
\hspace{1em}\Cb{h-method} & Heading (e.g, \ExampleQuote{Method:}) & 162\\
\hspace{1em}\Cb{a-method} & \Cb{method} announcement (e.g., \ExampleQuote{A series of experiments is conducted.}) & 33\\
\addlinespace[0.3em]
\multicolumn{3}{l}{\textbf{\emph{Result} (\texttt{r})}}\\
\hspace{1em}\Cb{result} & Immediate, empirical outcome of the study & 1494\\
\hspace{1em}\Cb{h-result} & Heading (e.g, \ExampleQuote{Results:}) & 160\\
\hspace{1em}\Cb{a-result} & \Cb{result} announcement (e.g., \ExampleQuote{We identify key features.}) & 105\\
\hspace{1em}\Cb{claim} & Non-empirical would-be \Cb{result} statement (e.g., \ExampleQuote{This enables highly accurate code completion.}) & 29\\
\addlinespace[0.3em]
\multicolumn{3}{l}{\textbf{\emph{Summary} (\texttt{s})}}\\
\hspace{1em}\Cb{summary} & Summarization of results, but no new information. & 79\\
\addlinespace[0.3em]
\multicolumn{3}{l}{\textbf{\emph{Conclusion} (\texttt{c})}}\\
\hspace{1em}\Cb{conclusion} & Take-home message, less specific than one or more \Cb{result}s. & 369\\
\hspace{1em}\Cb{h-conclusion} & Heading (e.g, \ExampleQuote{Conclusions:}) & 154\\
\hspace{1em}\Cb{a-conclusion} & \Cb{conclusion} announcement (e.g., \ExampleQuote{We summarize recommendations.}) & 42\\
\addlinespace[0.3em]
\multicolumn{3}{l}{\textbf{\emph{Outlook} (\texttt{O})}}\\
\hspace{1em}\Cb{fposs} & Research that is now possible. & 83\\
\hspace{1em}\Cb{fneed} & Future research that should be done. & 50\\
\hspace{1em}\Cb{a-fposs} & \Cb{fposs} announcement (e.g., \ExampleQuote{We propose future studies on the topic.}) & 36\\
\hspace{1em}\Cb{a-fneed} & \Cb{fneed} announcement (e.g., \ExampleQuote{Our findings emphasize the need for future research.}) & 4\\
\hspace{1em}\Cb{h-fwork} & Heading (e.g, \ExampleQuote{Future work:}) & 2\\
\midrule
\Cb{:i} & Informativeness gap (subjective assessment, see Section~\ref{subjective-codes}) & 599\\
\Cb{:u} & Understandability gap (subjective assessment, see Section~\ref{subjective-codes}) & 134\\
\Cb{ignorediff} & Inherent ambiguity (coders agree to disagree, see Section~\ref{ignorediff}) & 48\\
\bottomrule
\end{tabular}
\end{table*}

\end{knitrout}

\subsection{Training}\label{meth_training}

The training phase (internally called ``prestudy2'') served two purposes:
Finding/repairing deficiencies in the codebook,
and arriving at a joint interpretation of it across the four coders
(the four authors\footnote{Four additional people were involved in the training phase at some
  point, but they decided not to join the full study.}).
As can be seen in Figure~\ref{qabstracts_timeline_commits},
the training phase extended over almost half a year and triggered
the majority of the codebook improvements.

In the training phase, we perfected the mechanics of the coding process
(described in Section~\ref{meth_coding}) and generally formed as a research team.

\subsection{Sample}\label{meth_sample}

We decided not to aim for a broad selection of all software engineering research,
but rather concentrate on what is presumably the highest quality material:
The ICSE technical research track,
the three journals allowed for journal-first presentations at ICSE,
i.e., Empirical Software Engineering (\textit{EMSE}),
ACM Transactions on Software Engineering and Methodology (\textit{TOSEM}),
IEEE Transactions on Software Engineering (\textit{TSE}).
Since we expected to find that structured abstracts had better quality
than unstructured ones, we added a fifth venue that \emph{requires} structured abstracts:
Information and Software Technology (\textit{IST}, an Elsevier journal).
IST has published many very good systematic literature reviews and methods works,
but is not \emph{generally} considered a top-quality venue.

We wanted to draw a random sample of 100 articles per venue from the
2022 volumes, but found that TOSEM had published only 86 articles that year,
so we ended up at 486 articles initially.
Of those, a few later had to be removed because they were not research articles (often editorials).
Furthermore, we eventually did not need quite as much data for answering our
questions and stopped coding after 362 abstracts.\footnote{After abstracts
	were dropped, our sample is no longer perfectly balanced, with EMSE:73, ICSE:74, IST:71, TOSEM:71, TSE:73 abstracts.}
Still, ours is the largest manual study of abstracts we know of.

Volume downloading, sampling, and abstract extraction into publishable
and annotation-ready text files were all done automatically by the scripts
\Prg{retrievelit}\footnote{\url{https://github.com/serqco/retrievelit/}},
\Prg{select-sample}, and \Prg{prepare-sample} --- except
that the EMSE article format required manual cleansing.
All these tools were purpose-built for the present study.

\subsection{Coding Process}\label{meth_coding}

We coded each abstract twice, by so-called coders A~and~B.
Abstracts are held in text files in separate directories 
\Repo{\Prg{abstracts.A/}}{abstracts/abstracts.A} and 
\Repo{\Prg{abstracts.B/}}{abstracts/abstracts.B}.
Each sentence is followed by a line containing a pair of double curly braces \Prg{\{\{\}\}}
into which the coder would enter their codings,
such as \Prg{\{\{method,result:i2\}\}} (for a complex sentence that contains substantial
amounts of method information as well as results with two informativeness gaps).
We batched the coding in blocks of 8 abstracts each.
Coders picked and processed blocks based on their available time, resulting in different numbers of blocks
done by each author, between 13 blocks for Franz and 31 blocks for Lutz. 
The procedure, coordinated via git, is best explained by example,
which we do in the following two subsections.

\subsubsection{Coding}

\textbf{Step~1.} When Lloyd wanted to code a block of abstracts on 2023-05-26,
he found the next available block to be Block~17, which was already coded once by Lutz (``coder~A'').
Lloyd reserved his spot as ``coder~B'' in the coordination file 
\Repo{\Prg{sample-who-what.txt}}{abstracts/sample-who-what.txt}
and performed the coding.\\
\textbf{Step~2.} He then ran the 
\Repo{\Prg{check-codings}}{script/qscript/cmd/check_codings.py} 
script to test his codings against the codebook and corrected any mistakes, such as typos.\\
\textbf{Step~3.} He then ran 
\Repo{\Prg{compare-codings}}{script/qscript/cmd/compare_codings.py}
to compare his codings against Lutz'.
This script creates one \emph{report block} for each sentence where the codings
of coders A and B are not compatible.
We define codings to be compatible if they differ 
at most in the subjective suffixes \Cb{:i} and \Cb{:u}, but not in the codes
(and also not by more than one in the numbers of informativeness gaps and understandability gaps).
If Lloyd found a report block where Lutz' coding was obviously correct and his own
obviously wrong, he would simply correct his coding.\\
\textbf{Step~4.} He would then commit his coded abstracts into git.

\begin{lstlisting}[language={},
  caption={Example of a coded abstract: \Art{RosClaMad22}.},
  captionpos=t,
  label={lst:coding_example},
  breaklines=true,
  breakindent=0pt,
  moredelim={[s][keywordstyle]{\{\{}{\}\}}},  % Adds bold styling to our codes. Added for visual clarity
  keywordstyle=\ttfamily\bfseries\color{black},
  basicstyle=\scriptsize\color{abstractcolor}]
Empirical Effort and Schedule Estimation Models for Agile Processes in the US DoD.

Estimating the cost and schedule of agile software projects is critical at an early phase to establish baseline budgets and schedules for the selection of competitive bidders.
{{background}}
The challenge is that common agile sizing measures such as story points and user stories are not practical for early estimation as these are often reported after contract award in DoD.
{{gap}}
This study provides a set of effort and schedule estimation models for agile projects using a sizing measure that is available before proposal evaluation based on data from 36 DoD agile projects.
{{objective,method}}
The results suggest that initial software requirements, defined as the sum of functions and external interfaces, is an effective sizing measure for early estimation of effort and schedule of agile projects.
{{conclusion}}
The models' accuracy improves when application domain groups and peak staff are added as inputs.
{{conclusion}}
---
\end{lstlisting}

\subsubsection{Handling Disagreements}
\label{ignorediff}

\begin{figure*}[tbp]%
	\centering\fbox{\includegraphics[width=0.95\textwidth]{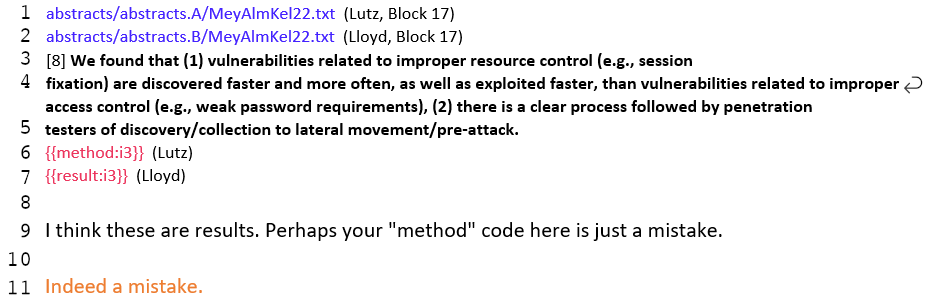}}%
	\vspace{-2mm}\caption{Excerpt from Lutz' response email during the disagreements handling
		for Block 17.
		Lines 1--7: Report block generated by the \Prg{compare-codings} script (\textbf{Step 3});
    line 9: text line from Lloyds first email (\textbf{Step 5});
    line 11: Lutz' response (\textbf{Step 6}).
		This one was a simple case;
		sometimes each coder provided several sentences of argumentation.
		Overall, Lloyd's first email contained report blocks for 9 disagreements
		in 4 abstracts, both typical numbers.}\label{email-MeyAlmKel22.png}%
\end{figure*}

\textbf{Step~5.} For the remaining (unresolved) report blocks, 
Lloyd would write an email to Lutz explaining his reasoning.\\
\textbf{Step~6.} Lutz would read through that email and categorize the report blocks into the following cases:
\begin{enumerate}[label=\alph*)]
  \item Lloyd's coding is obviously correct
  (and Lutz' own is a clerical error, as in Figure~\ref{email-MeyAlmKel22.png}) or
  Lutz prefers Lloyds coding.
  Lutz adjusts his coding accordingly.
  \item Lutz finds Lloyd's coding clearly incorrect (clerical error) or
  Lutz prefers his own coding.
  He would respond with an explanation of his reasoning and suggest that
  Lloyd either adjust his coding or add an \Cb{-ignorediff} marker to it.
  This suffix indicates two accepted alternative interpretations of the same sentence.%
  \footnote{It silences the 
    \Repo{\Prg{compare-codings}}{script/qscript/cmd/compare_codings.py}
    script for this particular report block, so that the coding difference is now officially accepted.}
  \item Lutz finds both codings equally acceptable.
  He would add an \Cb{-ignorediff} marker to his own and respond accordingly, explaining his reasoning.
\end{enumerate}
\textbf{Step~7.} Lutz would commit his corrected abstracts and send the response email.\\
\textbf{Step~8.} Lloyd would read the response email and usually act on it to finish the handling of
Block~17.
Only rarely would he disagree with something to a degree that would make
another round of emails necessary.

\subsubsection{Effect of this Procedure}

Most research involving content analysis codes most of their raw data only once,
then codes a random subsample a second time,
computes some coefficient of agreement,
reports it as a measure of good-enough coding quality,
and that's that.
In contrast, our above-described procedure has two effects:
\begin{enumerate}
	\item It maximizes the quality of the coding.
	  Very few clerical errors, if any, will have managed to escape our discussion process.
	  Besides, the definitions of all codes that are sometimes difficult to tell apart
	  have been refined until they were very mature.
	\item It finds all those spots in the sample where the abstracts are so convoluted
	  or strangely formulated that even our careful code definitions do not lead
	  to a canonical judgment.
	  These spots, which will be marked by a \Cb{-ignorediff} annotation in our data,
	  should clearly be considered to be badly written.
	  In this manner, our coding produces additional findings of a relevant type.
\end{enumerate}
Note that, technically speaking, our procedure also leads to a 100\% inter-coder agreement,
because even the cases of \Cb{-ignorediff} indicate that
the coders agree on multiple plausible interpretations of one sentence.

\subsection{Statistical Evaluation}\label{statisticalevaluation}

The difficult part of our statistical evaluation is asking the right questions;
our data provides a lot of possibilities.
In contrast, the actual statistical techniques are simple and straightforward:
Mostly tabulations of counts or percentages, bar plots, and box plots.
We define two binary properties of abstracts: 
A \emph{complete} abstract contains all of the basic elements described in 
Section~\ref{meth_basicparts}.
A \emph{proper} abstract is complete and does not have any of the deficiencies
described in Section~\ref{results} as making an abstract \emph{improper}.

Since we coded each abstract twice, we get potentially conflicting assessments regarding these binary properties.
We calculate and report percentages across all \emph{codings}
($n_C\,{=}\,724$, which each count as a ``half'' abstract),
as well as lenient and conservative percentages per \emph{abstract}
($n_A\,{=}\,362$),
for which either both or just one coder are necessary to attest a deficiency, respectively.
We report these numbers with $\pm$ in the text and as error bars in the plots.
When we write
\Pseudoquote{29\%$\pm$0.6\% of the abstracts are complete},
we mean that 29\% of the 724 abstract codings qualify
the corresponding abstract as \emph{complete},
and that for 28\% of the 362 abstracts both coders agreed,
while at least one coder thought so for 30\% of them.

We mostly refrain from performing significance tests or computing confidence intervals,
because most analyses are exploratory, not driven by specific expectations or theories.
The one exception from this rule is the comparison of
structured versus unstructured abstracts (Section~\ref{structuredgood}).

Most phenomena we report are gradual by nature.
In this spirit, we use the following verbal terms for frequencies:
rare (less than 5\%),
not rare (5-20\%),
common (20-35\%),
frequent (35-50\%),
dominant (over 50\%).

\subsection{Interpretation}

Our interpretation of the measurements is driven by our research interest and
guided by our expertise as software engineering researchers who read abstracts.
Whenever we mark a phenomenon as a weakness, we will provide a justification from that angle
and provide an example to allow the reader to relate to the justification.

\subsection{Readability Metric}

For evaluating general readability at a purely language level,
we use the Flesch-Kincaid ``reading ease'' readability score \cite{KinFisRog75}.
It is a validated and widely used metric that judges readability of English text
based only on the number of words per sentence and the number of
syllables per word.
Values under 30 represent graduate-level difficulty,
under 10 extreme difficulty for native speakers.
Considering that most community members are not native speakers of English,
we judge 20 to 30 to be a range suitable for researcher audiences and so
we call anything over 30 ``good'',
20 to 30 ``normal'', and
under 20 ``improper''.

\subsection{Use of Examples}

We will use examples from real abstracts from our sample to illustrate
some of our statements.
We identify their source by a citation key
formed from three letters each of the first three authors' names.
For example article 1 in Block 1 in our study was written by
Fregnan, Petrulio, Di Geronimo, and Bacchelli
and is thus identified as \Art{FrePetGer22}.

We select those examples for the clarity of the phenomenon in question,
not for the abstract's quality.
For the source of every positive example there are others that are
as good or better.
For the source of every negative example there are others that are
as bad or worse.
Since the use of an example is not about the article it stems from,
we do not include those articles in our references list.

%========================================================================
\section{Results}
\label{results}

\subsection{Length and Readability}
\label{result-len-read}

Typical abstracts (the middle half) are 190 to 280 words long
and 8 to 14 sentences long.
Structured abstracts tend to be longer than unstructured ones.
\ifwithappendix{See Figures \ref{boxplots_words} to \ref{boxplots_avg_wordlength} for details.}
Some venues have official abstract length limits:
150--250 words for EMSE,
maximum 300 for IST, and
recommended maximum 250 for TSE.
These limits are disobeyed by more than a quarter of all articles for each of these three venues.

For readability, see Figure~\ref{boxplots_fkscore}:
only about 13\% of all abstracts have good readability (>30),
about 35\% have normal readability (20-30),
and 52\% are improper (<20) ---
more than every fifth abstract (21\%) even has a score below 10.
ICSE tends to be better than the other venues,
TOSEM is worse.

\Plot{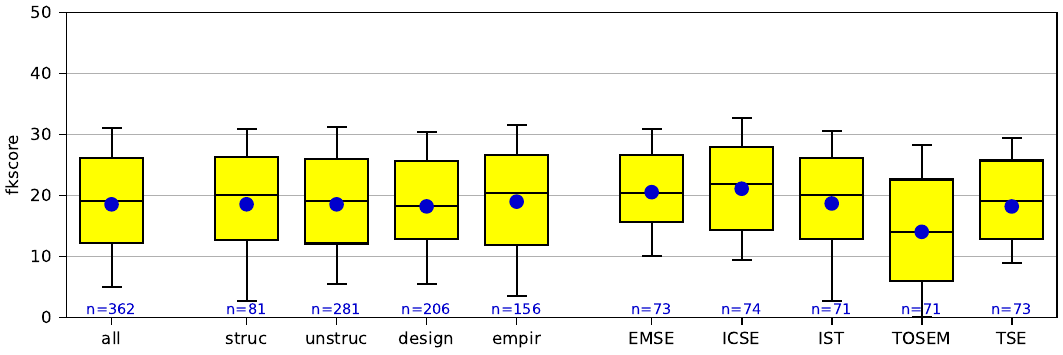}{%
	Flesch-Kincaid `reading ease' readability score, higher is better.
    Values under 50 are considered difficult to read (college-level material);
    under 30: very difficult to read (graduate level, acceptable for abstracts);
	  under 10: extremely difficult to read (overly difficult).
    Differences between subgroups are modest.
    ICSE is best (mean 21),
    TOSEM is worst (mean 14).
    The abstract of the present article has a score of~33.
    Structured abstracts are just as difficult as nonstructured ones
    if one ignores the ``Methods:'' etc. headings as we did for the readability analysis.
}

\subsection{Design Articles vs. Empirical Articles}

We described the idea of the \Cb{design} code in Section~\ref{design}.
But many empirical works involve some artifact design as well,
so how is the discrimination made?
That depends on how the authors phrase their goal in their
\Cb{objective} statement.
Consider the following goal statements:
\Quote{In this paper, we propose AI-based automation for the
completeness checking of privacy policies.} (from \Art{AmaAbuTor22}).
\Quote{This study aims to investigate the urgency and importance of
reproducibility and replicability for DL studies on SE tasks.} (from \Art{LiuGaoXia22}).

The first puts the artifact at the center, so this is a design article.
The second puts empirical results at the center, so this is an empirical article
and the \Cb{design} code will \emph{not} subsequently be used.
Artifact design discussions are then usually coded as \Cb{method} instead.
In mixed cases, which are rare, the two coders would decide which aspect has more weight.

In design articles, a large fraction of the abstract will be devoted to
describing design considerations for that artifact -- typically 
14\% to 33\% of the words in our data.
The empirical study, which commonly exists as well, then usually has a mere
supporting role (validating the claims made in the artifact discussion)
and is correspondingly given less space:
5\%--13\%
(versus 11\%--24\% for empirical works) 
for description of empirical method,
10\%--22\%
(versus 16\%--33\%)
for description of empirical results.
Design article abstracts are barely longer than empirical article abstracts.

\subsection{The Abstracts Archetype}\label{archetype}

Compared to the content structure assumed by the usual formats of
structured abstracts, our codebook is more fine-grained;
the \Cb{design} code is but one example.

During our sensemaking process, we had a number of insights regarding
how a well-written abstract ``ticks'', which we eventually distilled into
the following template, which we call the 
\emph{SERQco software engineering abstracts archetype}\footnote{SERQco is the
  Software Engineering Research Quality Coalition}
(see also Figure~\ref{abstracts_archetype_fig}):

\begin{enumerate}
\item An abstract consists of three parts, in this order:
   \emph{Introduction}, \emph{Study Description}, and \emph{Outlook}.
\item Two turning points connect the three parts:\\
   a) A statement of the study goals (\Cb{objective}) connects \emph{Introduction}
      to \emph{Study Description}.\\
   b) A generalizing statement (``take-home message'', \Cb{conclusion})
     connects \emph{Study Description} to \emph{Outlook}.
\item The \emph{Introduction} first introduces the topic area of the study and what is known (\Cb{background})
   and then may or may not point out a gap in knowledge (\Cb{gap}).
\item For an empirical article, the \emph{Study Description} begins with
   method description (\Cb{method}), followed by results description (\Cb{result}).
   Sometimes, this sequence occurs twice in a row, very rarely more.
\item For a design article, design description (\Cb{design}, see below) precedes the structure
   described in the previous item.
% \item Occasionally (but infrequently), \emph{Study Description} will end with a study summary (\Cb{summary}).
\item After the \Cb{conclusion}, the \emph{Outlook} talks about future research and states
   what could now be done (\Cb{fposs}, for future possibilities),
   what should now be done (\Cb{fneed}),
   what the authors themselves intend to do (\Cb{fwork}), or
   what is still not known (\Cb{fgap}).
   Several statements of each type may occur (including none), in no particular order.
   In most cases, the space devoted to \emph{Outlook} would be more informative if spent on
   \emph{Study Description}.
\end{enumerate}

The archetype describes all variants of abstracts that have a natural train of thought.
It is an engineering-specific generalization of the well-known IMRAD structure \cite{SolPer04}.
Deviations from the archetype will tend to lead to a less easily understandable abstract.

\PlotWidth{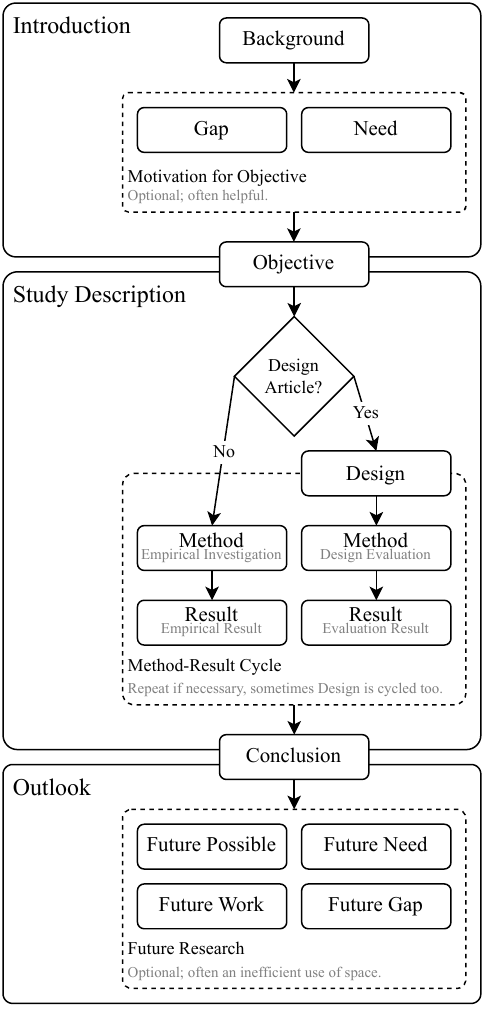}{.8}{%
	A visual representation of the SERQco abstracts archetype. 
	Note that this figure only describes the ``natural train of thought'' and not all possible abstract forms 
	that we coded in this study. 
	This is the recommended structure.}

\subsection{How Not To: Inefficient Allocation of Space}
\label{inefficient_allocation}

If one reads the abstracts in our sample, one can hardly help notice that some of them 
spend a lot of space on \Cb{background}, although its only purpose is to situate
the objective and make it understandable.
For example, \Art{FirFirRos22} spends 66\% of the abstract
for explaining their rather niche application domain in some detail
rather than just naming it and letting interested readers look up the rest in the article body.

\Plotwide{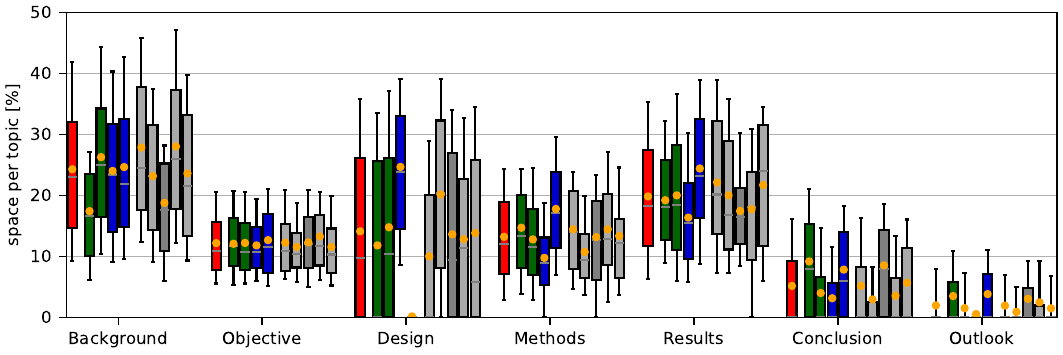}{%
	Per-topic distribution of the amount of space used for that topic.
	These ``topics'' are groupings of related codes
  (see also Table~\ref{tab:codes});
  e.g., Outlook stands for the union of \Cb{fposs}, \Cb{fneed},
	and all corresponding \Cb{a-*} and \Cb{h-*} codes.\\
	\Describeboxplots
	\Describegroups}

% This figure we shifted up to get a reasonable layout, see below:
\Plotwide{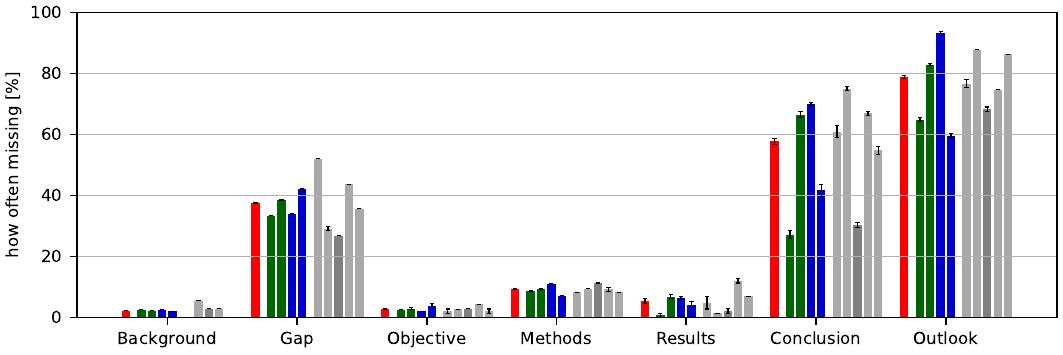}{%
	How often is a topic not present at all in an abstract?
  These ``topics'' are groupings of related codes (see Table~\ref{tab:codes}).\\
	\Describegroups\ 
  Error bars are due to ambiguous formulations (see Section~\ref{ignorediff}).}

\Plot{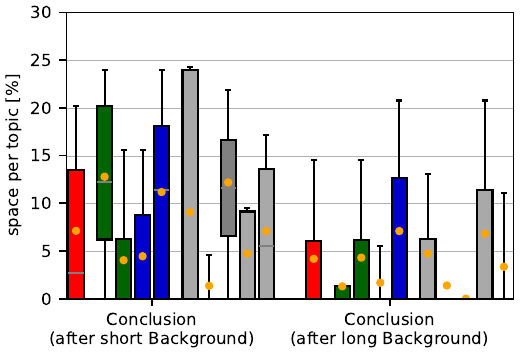}{%
	Comparison of the amounts of remaining space devoted to the conclusion for abstracts
	with a rather short background section (lowest quarter) vs those
	with a long one (highest quarter).
	The latter conclusions are shorter even though the indicated percentage 
	pertains to the part of the abstract after the background only.\\
	\Describegroups}

As we can see in the Background group of boxplots in Figure~\ref{box_xletgroups_topicfractions},
all venues that do not require structured abstracts have at least a quarter of articles 
that spend over 30\% of the abstract space on background. 
As we can see in Figure~\ref{box_xletgroups_conclusionfractions},
this leads to deficiencies later on:
The conclusion is the potentially most useful part of the abstract, the take-home message,
but with a long background section, it tends to become pronouncedly shorter.

Background lengths are more benign for structured abstracts.

\subsection{How Not To: Missing Elements}
\label{missing_elements}

Given the archetype, one way to approach an analysis of abstracts quality is to ask
how often key parts of an abstract are missing entirely.
This is shown in Figure~\ref{zerofractionbar_xletgroups_topicmissingfractions}.

% Logically, Figure zerofractionbar_xletgroups_topicmissingfractions belongs here
% We include it earlier for layout reasons, as otherwise figures appear too late.

% rare:     <5%
% not rare: 5-20%
% common:   20-35%
% frequent: 35-50%
% dominant: 50%

The \Cb{gap} and \emph{Outlook} parts \Cb{fposs}, \Cb{fneed}, etc. are clearly optional,
so it is not a problem that their absence is
frequent (40\%) and
dominant (79\%), respectively.
\Cb{background} and \Cb{objective} are rarely missing
(2\% and 3\%).

Missing \Cb{method} and \Cb{result} are not rare
(11\% and 10\%),
which we find alarming.
Both are always present in our structured abstracts.

The shocking part of this analysis is \Cb{conclusion}, which ought to be present
as the key take-home message in any abstract, but is in fact missing in more than
half of all (62\%).\footnote{This value is higher
  than the 'all' bar in Figure~\ref{zerofractionbar_xletgroups_topicmissingfractions},
  because that bar shrinks when an \Cb{h-conclusion} or \Cb{a-conclusion} appears,
  but those do not serve the \Cb{conclusion} purpose.}
The situation is better for structured abstracts, but missing \Cb{conclusion}
are still common here
(29\%).%
\footnote{The prompt of having to write something after a \ExampleQuote{Conclusion:} keyword
  cannot guarantee an actual \Cb{conclusion}:
  Dominantly, the statement authors put there is actually
  a \Cb{result} (an immediate empirical result),
  a \Cb{summary} (a result repetition) or still something else.}

Overall, only 29\%$\pm$0.6\% of all abstracts are \emph{complete}
in the sense that they contain all basic elements
\Cb{background} (or \Cb{gap}), \Cb{objective}, \Cb{method}, \Cb{result}, and \Cb{conclusion}.

\subsection{How Not To: Confusing abstracts}
\label{confusing}

Any deviation from the natural train of thought of the abstracts archetype
will tend to increase the reader's cognitive load.
If this happens, we call the abstract's train of thought convoluted.

For investigating this issue, we map each abstract's structure to a sequence of letters,
where each letter stands for a contiguous stretch of sentences in the abstract
that have the same topic (see Table~\ref{tab:codes}).\footnote{However, if a, say, 
  \Cb{h-conclusion} is followed by, say, a \Cb{summary} (which is a common style), 
  this would result in a topicletter string that looks more complicated than the abstract actually reads.
  We find this misleading and hence exclude \Cb{h-*} codes from the topicletter string.}
For instance, an abstract that follows a minimal incarnation of the archetype
for an empirical article would be encoded as \strct{bomrc}, which stands for
the topic sequence \textit{background, objective, methods, results, conclusion}.
Non-minimal incarnations allow for many different structures but
even complex structures (with long strings) \emph{can} be easy to read.
Other structures, however, can be problematic.
The full list of abstracts structures is too long to discuss it here:
It has 117 entries for empirical articles and
125 for design articles.

We have found no simple criterion to reliably diagnose which of these to consider convoluted
and which not, so this section will only provide examples and does not quantify
how many abstracts are actually convoluted.
Furthermore, even abstracts that \emph{do} conform to the archetype
can be confusing for semantic reasons.
We will include examples of this type as well. 

Abstract structures with more than two instances are shown in 
Figure~\ref{ab_topicstructure_freqs_empir} for empirical articles and
Figure~\ref{ab_topicstructure_freqs_design} for design articles.
(Fractional frequencies such as $2.5$ mean that the two coders did not agree.)

\Plot{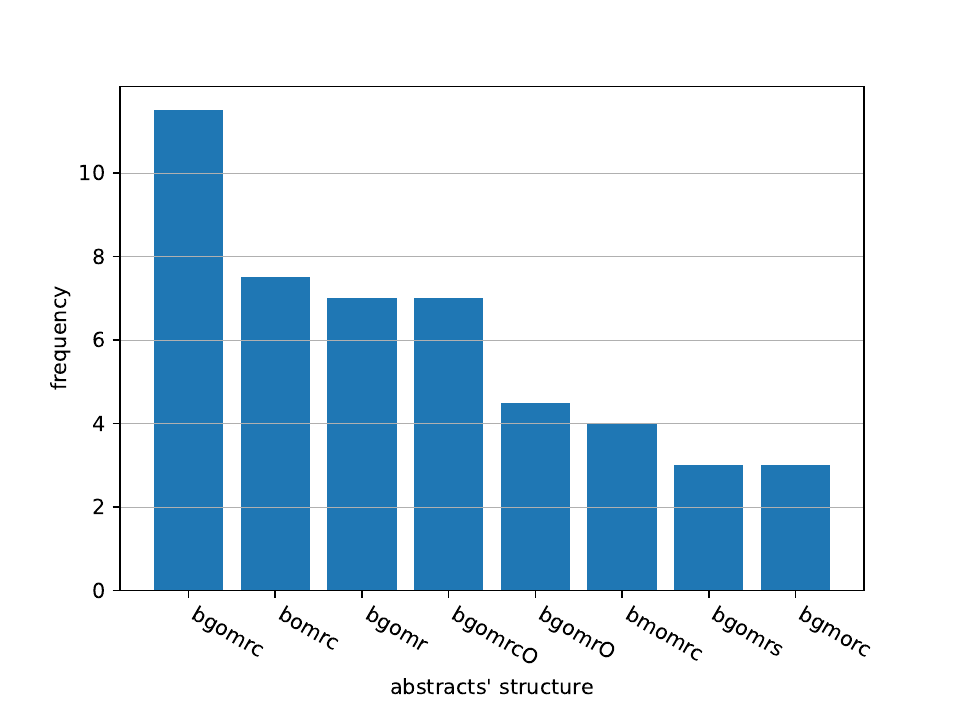}{%
  The frequency of different trains-of-thought in the abstract for empirical articles.
  The label is a string of stretch-code characters:
  b\mbox{-}ackground, g\mbox{-}ap, o\mbox{-}bjective, d\mbox{-}esign, m\mbox{-}ethod, r\mbox{-}esult,
  s\mbox{-}ummary, c\mbox{-}onclusion, O\mbox{-}utlook.}
\Plot{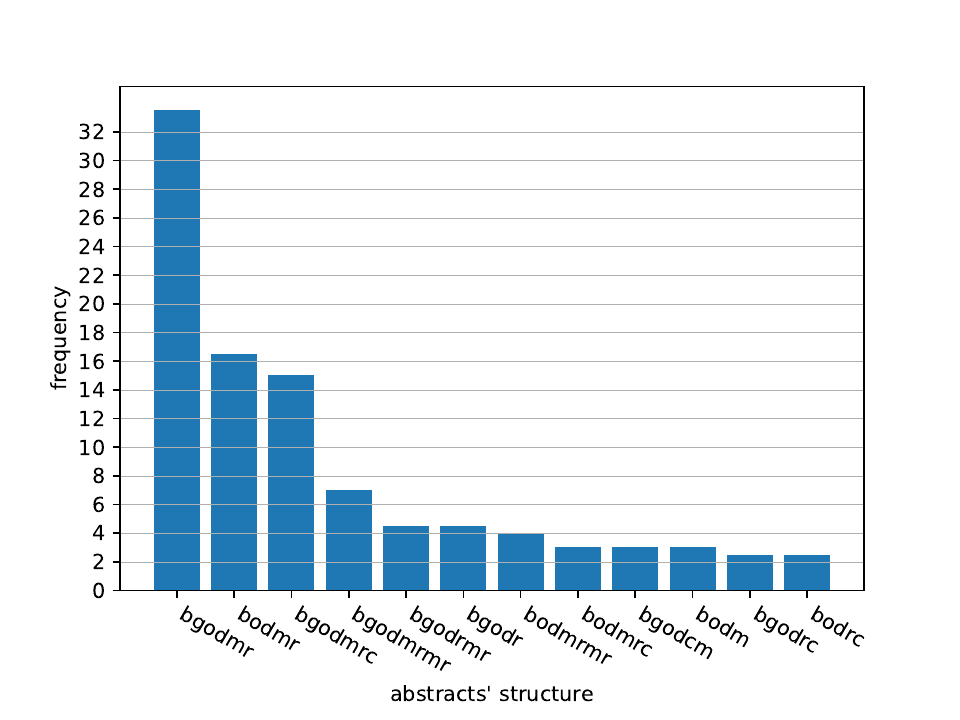}{%
  The frequency of different trains-of-thought in the abstract for design articles.
  Same characters as before, except that d\mbox{-}esign can now in fact occur.}

\subsubsection{Structured Abstracts}

Structured abstracts help reduce confusion in two ways:
They standardize the order in which ideas are presented, so readers know what to expect when.
And their subheadings announce specifically what is to come next, which avoids many ambiguities.

Today's conventions for structured abstracts may be a bit restrictive
(e.g. by not accommodating the useful \strct{mrmr} substructure),
but they do indeed result in more orderly abstracts:
There are only 31 different abstract structures
for the 84 structured abstracts of empirical articles,
but 56 different abstract structures on average
for a random sample of the same size of the non-structured abstracts of empirical articles.
% \footnote{95\%~CI
%   [round(n_distinct_empir_nonstruc_ci[1]), round(n_distinct_empir_nonstruc_ci[2])].}

The orderliness of a structured abstract may not help, however, if the supposed structure is broken.
We have seen only few cases of this for empirical articles, 
but once in a while some piece of information appears not where it should
(such as \Cb{Background} in the \Sah{Objective} section in \Art{LiaGaoXia22})
or authors appear to have added the subheadings to an already written non-structured abstract
without rewriting it, so that lots of sections contain information of the wrong type
(as in \Art{SelOunSai22}).

The latter of these occurs in design articles as well (as in \Art{GeFanQia22}).
The former is even unavoidable for a design article, as no proper place for design information
exists in today's conventional structured abstract section list.

% #confusing LiaGaoXia22(e, struc, background in h-objective section)
% #confusing SelOunSai22(e, struc, misplaced info in several sections),
% #confusing GeFanQia22(d+struc with bad fit),

\subsubsection{Empirical Articles}

As we see in Figure~\ref{ab_topicstructure_freqs_empir},
there are only 8 % #nonknitr
among the 31 different abstract structures of empirical works
that occur more than twice.
Of these, bars 5, 6, 7 show abstracts with a non-archetypical train of thought:
They mix \Cb{method} and \Cb{objective} 
(\strct{bmomrc} and \strct{bgmorc})
% #nongood \strct{bmomrc}: \Art{ForStoZim22}, \Art{NolShaGao22}, \Art{PapGroSub22}, \Art{RahImtSto22}, 
% #nongood \strct{bgmorc}: \Art{LiPenXia22}, \Art{MasHodBli22}, \Art{WanZhaZha22})
or have a trailing \Cb{result} after the \Cb{conclusion} 
(\strct{bgomrcr}).

When we annotated the abstracts, we kept notes on remarkable cases of all kinds.
Combing through those notes, we can look for patterns of actually confusing abstracts.
For the empirical works, 
there are some abstracts with information that is partially out-of-place or mis-shaped 
and will increase cognitive load for readers, 
but they are diverse and hardly worth calling a common pattern; for example:
\Cb{method} before \Cb{objective} (\Art{PreMohVil22}),  % #nongood
additional \Cb{background} late in the abstract (\Art{UddGuéKho22}),  % #nongood
a sudden second objective in the very last sentence (\Art{DanPlaHer22}),  % #nongood
an objective that sounds like from a design work (\Art{GaoZhuZha22}),  % #nongood
repeating (in different words) the objective where a conclusion should be (\Art{AhmMerBah22}),  % #nongood
or mixing present tense and past tense such that results sound 
as if they were conclusions (\Art{ValHunFig22}).  % #nongood

% #confusing AhmMerBah22(e, repeats o in different words instead of a conclusion)
% #confusing DanPlaHer22(e, new 2nd o in last sentence),
% #confusing GaoZhuZha22(e, but objective sounds like d),
% #confusing PreMohVil22(e, method before objective),
% #confusing UddGuéKho22(e, background later in abstract to explain study history),
% #confusing ValHunFig22(e, mixing present+past tense so that results sound like concls),
% #notused   MaMocZar22(e/d, struc, conclusion in objective section)

\subsubsection{Design Articles}\label{structure-design-articles}

As we see in Figure~\ref{ab_topicstructure_freqs_design},
many of the common structures of design works lack some expected parts, 
most often a conclusion, which is missing for all bars except numbers 3, 8, and 10.
Yet in terms of the \emph{order} of things, only two structures deviate from the archetype:
\strct{bodm} (bar 9) and 
\strct{bgodcm} (bar 10),
both of which curiously end with \Cb{method} information.
Having multiple \strct{mr} pairs is more common for design works than it is for empirical works.

As for truly confusing abstracts, the most conspicuous pattern is an unclear answer to the
most fundamental question: Is this a design work or an empirical work?
This issue occurs in various forms:
a clear design \Cb{objective}, but a solely empirical \Cb{conclusion} (\Art{CorRweFra22}),  % #nongood
the \Cb{objective} is phrased like for an empirical work, 
but the \Cb{conclusion} makes clear it is a design work (\Art{BocSchApe22}, \Art{SuFanChe22}),  % #nongood
a clear design \Cb{objective}, but not a single \Cb{design} statement follows (\Art{FirFirRos22}, \Art{HerFerGal22}),  % #nongood
there are two \Cb{objective} statements, placed non-continguously, and the first is clearly empirical (\Art{MadNagBir22}),  % #nongood
or many statements throughout have an unclear role (\Art{NaeAla22}).  % #nongood

% #confusing CorRweFra22(d objective, e conclusion),
% #confusing FirFirRos22(d objective, but no d stmts),
% #confusing BocSchApe22(e->d, obj sounds like e, but the conclusion makes it clearly d),
% #confusing HerFerGal22(d with no design, only a-design),
% #confusing MadNagBir22(d, first of two non-contiguous objectives is clearly e),
% #confusing SuFanChe22(d, but objective can also be read like e),
% #confusing NaeAla22(d, many sentences have unclear role),

Of course, cases of misplaced information occur in design works as well, e.g.
a \Cb{result} in the middle of a very long \Cb{background} section (\Art{NiuWuNie22}),  % #nongood!!!
\Cb{background} placed behind \Cb{objective}, which makes it sound a lot like a \Cb{result} (\Art{LuLiLiu22}),  % #nongood!!!
\Cb{method} and \Cb{result} placed after \Cb{h-conclusion} (\Art{HumKho22}, \Art{ImrDam22}).  % #nongood

% #confusing NiuWuNie22(d, result in the very long background section),
% #confusing LuLiLiu22(d, background after objective risks confusion with result)
% #confusing HumKho22(d, struc, with m and r after h-conclusion),
% #confusing ImrDam22(d, struc, again with m and r after h-conclusion),

Other peculiarities appear in design works that we have not seen in empirical works, e.g.
an \Cb{objective} that is difficult to recognize as such (\Art{TiaLiPia22}),  % #nongood!!!
formulating a \Cb{conclusion} without ever presenting a \Cb{result} (\Art{LiaHanLi22}),  % #nongood!!!
or a sequence of \Cb{objective}, \Cb{objective}, \Cb{design}+\Cb{result} phrased as a first{\ldots}then{\ldots}finally structure (\Art{GueLarChe22}).  % #nongood

% #confusing LiaHanLi22(d, conclusion without results),
% #confusing GueLarChe22(d, obj,obj,design+result phrased as first-then-finally),
% #confusing TiaLiPia22(d, objective is difficult to recognize as such),

\subsection{How Not To: Uninformative Formulations}\label{uninformative}

Another way of reducing the usefulness of an abstract is using formulations that
fail to provide information that, at this point, would be useful to the reader and
could be provided using only very few additional words.

Our investigation has identified and then quantified two types
of such lacks of informativeness.
We call them informativeness gaps and announcements, respectively.

\subsubsection{Informativeness Gaps}\label{igaps}

Look at the following result statement:
\Quote{random sampling is rare} \Art{BalRal22}.  % #nongood
If at this point the reader expects the work contains something more
concrete than \Quote{rare}, this is an informativeness gap.
And indeed the article in question contains this information in its
Table 4 and therefore could and should have said
\Pseudoquote{random sampling is rare \textbf{(8\% of cases)}}.

Informativeness gaps (coded \Cb{:i}, see Section~\ref{subjective-codes})
appear mostly
in \Cb{result} statements (83\% of the gaps)
or \Cb{method} statements (12\% of the gaps).
Most (if not all) of them could be filled by a number.
They tend to cluster,
like in \Art{ShiBiaBri22}:  % #nongood
\Quote{The results show that PRINS can process large logs much faster
  than a publicly available and well-known state-of-the-art tool,
  without significantly compromising the accuracy of inferred models.}
This sentence has three informativeness gaps.
A better formulation could have been:
\Pseudoquote{The results show that PRINS can often process logs \textbf{an order of magnitude} faster
  than the well-known state-of-the-art tool \textbf{MINT},
  but \textbf{never lost more than 7 percentage points} of balanced accuracy.}

More than half of all abstracts (55\%$\pm$11\%) have an informativeness gap
(making them \emph{improper})
and 18\%$\pm$5\% have three or more.
See the leftmost two groups of Figure~\ref{nonzerofractionbar_xletgroups_missinginfofractions}
for details.

\Plotwide{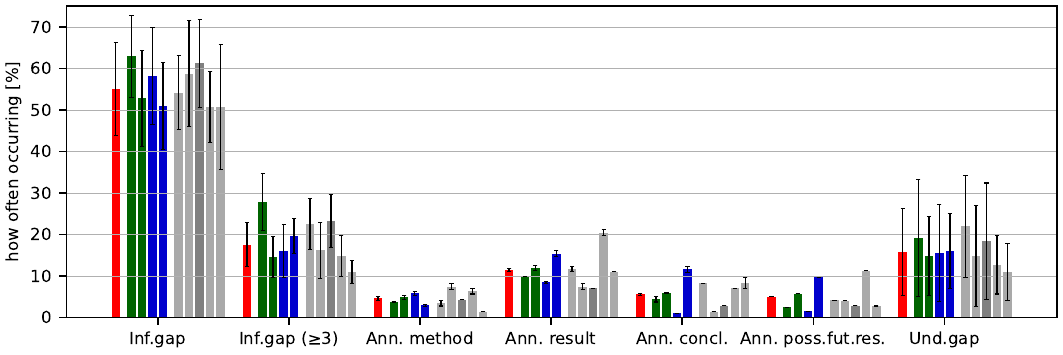}{%
	What fraction of abstracts has the following uninformative types of formulations?\\
  \Describegroups\
  The groups are:
	One-or-more or three-or-more informativeness gaps (i.e., missed opportunities for being more specific);
	some sentence that only announces (instead of describing) a method, result, conclusion, or
	possible future research;
  one or more understandability gaps.
  Error bars in the \emph{Inf.gap} and \emph{Und.gap} groups are due to differing coder perceptions
  of informativeness and understandability gaps (see Section~\ref{subjective-codes});
  error bars in the \emph{Ann.} groups are due to ambiguous formulations (see Section~\ref{ignorediff}).\\
	Informativeness gaps are epidemic, worse(!) for structured abstracts (1st green bar).
	Announcing is generally not rare, in particular for results, and tends to be less pronounced
	at IST (3rd gray bar).
	Not explaining key terms is not rare and worst at EMSE (1st gray bar).\\}

\subsubsection{Announcements}\label{announcements}

Occasionally, a sentence will not merely miss to report some specific piece of information
but rather fail to provide any useful information at all and merely hint at
information to be found in the article body.
We call such a sentence an announcement
(coded \Cb{a-*}, see Section~\ref{announcement-codes}).

Example (\Cb{a-method} from \Art{BesMarBos22}):  % #nongood
\Quote{As a second step, this study sets out to specifically
  provide a detailed assessment of additional and in-depth analysis of technical debt management strategies based
  on an encouraging mindset and attitude from both managers and technical roles to understand how, when and by
  whom such strategies are adopted in practice.} 
Here is what the sentence could have said instead:
\Pseudoquote{We then surveyed 26 managers and 46 technical people, followed by clarifying interviews
  with 4 managers and 2 developers in order to understand how the managers perceive how they are encouraging
  developers to manage technical debt and how the developers perceive the encouragement they receive.}
% based on BasMarBos Sections 3.2.1, 3.2.3, Table 2

Example (\Cb{a-result} from \Art{FliChaDye22}):  % #nongood
\Quote{Finally we provide guidelines/best practices for researchers utilizing time-based data
from Git repositories.} 
This should have been a result statement such as the following:
\Pseudoquote{We provide 6 guidelines. For instance, cutting off all commits before 2014
  will get rid of about 98\% of all bad commits.}

Announcements are a waste of space and a nuisance for the reader,
yet 24\% $\pm$0.6\%
of all abstracts have at least one and we consider those \emph{improper}.
See groups three to six of Figure~\ref{nonzerofractionbar_xletgroups_missinginfofractions}
for details.

\subsection{How Not To: Undefined Important Terms}\label{ugaps}

It is normal that the reader of an abstract has only a fuzzy understanding
of what certain terms in the abstract mean.
In a good abstract, the \emph{approximate} meaning of a statement using such a term
still comes across.
Sometimes, however, this is not the case:
The uncertainty regarding the meaning of the term weighs so much
that the sentence containing it becomes incomprehensible.
We call such a term use an understandability gap (coded \Cb{:u}, see Section~\ref{subjective-codes})
and consider such abstracts to be \emph{improper}.

Here is an example from \Art{HamMetQas22}:  % #nongood
\Quote{The results of evaluating the generality of the iContractML 2.0 reference model show
  that it is 91.7\% lucid and 72.2\% laconic.} 
Neither of the terms ``lucid'' or ``laconic'' have been introduced before, so that
this sentence has two understandability gaps and
an average reader is not able to make sense of this important statement
which represents half of the work's results.

See the rightmost group of Figure~\ref{nonzerofractionbar_xletgroups_missinginfofractions}
for how frequent this is:
About 16\%$\pm$10\% of all abstracts have one or more understandability gaps.
The issue is most pronounced at EMSE.

\subsection{How Not To: Ambiguous Formulations}\label{ambiguity}

Codings with an \Cb{-ignorediff} marker indicate cases where the coders could not agree
despite discussion (see Section~\ref{ignorediff}):
the respective sentence (or sentence part) is so highly ambiguous that
more than one role for it is plausible.
Obviously, such ambiguous formulations do not represent good abstract writing
and we consider such abstracts to be \emph{improper}.

In our data, we found 48 such cases overall, spread over 39 different abstracts,
so that 11\% of all abstracts have one or more such ambiguous sentences.
% A total of n_igdf_c different codes are involved.
% The most frequent 5 of them cover format_percent(r_igdf_cocodes_top5) of the total:
% igdf_cocodes_top5$code.

\subsection{How Not To: Summary}\label{hownotto_summary}

Summing up, a proper abstract should be \emph{complete}, i.e., it should provide at least 
a minimal amount of information of types  \Cb{background} (\Cb{or gap}),
\Cb{objective}, \Cb{methods}, \Cb{results}, and \Cb{conclusion} --- the
canonical IMRAD structure.

However, as we see in Figure~\ref{nonzerofractionbar_xletgroups_totalqualityfractions},
a majority of abstracts fails this very basic quality criterion, a depressing result.

\Plot{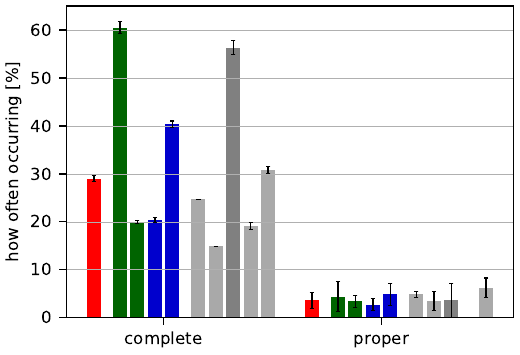}{%
	What fraction of abstracts is \emph{complete} in the sense of 
	Sections~\ref{statisticalevaluation} and \ref{hownotto_summary}?
	What fraction of abstracts is \emph{proper} in the sense of 
	Section~\ref{statisticalevaluation}?\\
  \Describegroups\
  Error bars for \emph{completeness} are due to ambiguous formulations (see Section~\ref{ignorediff}),
  error bars for \emph{properness} are due to differing coder perceptions of informativeness and understandability gaps
  (see Section~\ref{subjective-codes}).\\
  Only 29\%$\pm$0.6\% of abstracts are \emph{complete}. 
	Only about 20\% of unstructured abstracts and
  21\% of design article abstracts are complete.
	Structured abstracts are much better and therefore all four supposedly top-quality venues
	are beat by the much lower-regarded IST (3rd gray bar) which requires structured abstracts.\\
	Only 4\%$\pm$1.7\% of abstracts are \emph{proper}.
	Here, too, structured abstracts are better than unstructured ones.
	Our TOSEM sample has not a single proper abstract.
}

Our moderately stricter criterion of being \emph{proper} involves 
being \emph{complete}, having acceptable Flesch-Kincaid readability (>20),
and having neither informativeness gaps nor understandability gaps nor highly ambiguous sentences.
Given this list of criteria, improper abstracts will happen from time to time
even for careful authors.
Nevertheless, we believe that a majority of abstracts could and should be \emph{proper} in this sense.
Yet what we find is that only 4\%$\pm$1.7\% of them indeed are.
% TOSEM and ICSE are still worse than that.
We expected to find many problems in abstracts but still find this outcome astonishingly bad. 

\begin{knitrout}\scriptsize
\definecolor{shadecolor}{rgb}{0.969, 0.969, 0.969}\color{fgcolor}\begin{table}
\centering
\caption{\label{tab:overview}For each coding-based quality criterion,
      there is a possibility for disagreement between the
      two codings: For a negative criterion (italics),
      the lower value denotes the number of abstracts for
      which \emph{both} codings indicate a negative evaluation,
      the higher value denotes the number of abstracts for
      which \emph{at least one} coding indicates a negative
      evaluation.}
\centering
\begin{tabular}[t]{lr@{\hspace{2pt}}rr@{\hspace{2pt}}r}
\toprule
\multicolumn{1}{c}{\textbf{Criterion}} & \multicolumn{2}{c}{\textbf{Count}} & \multicolumn{2}{c}{\textbf{Ratio}} \\
\cmidrule(l{3pt}r{3pt}){1-1} \cmidrule(l{3pt}r{3pt}){2-3} \cmidrule(l{3pt}r{3pt}){4-5}
\addlinespace[0.3em]
\multicolumn{3}{l}{\textbf{Readability
    (Section \ref{result-len-read})}}\\
\hspace{1em}good ($fk > 30$) & 47~~ &  & 13\% & \\
\hspace{1em}acceptable ($fk \in [20, 30]$) & 126~~ &  & 35\% & \\
\em{\hspace{1em}improper ($fk < 20$)} & \em{189~~} & \em{} & \em{52\%} & \em{}\\
\addlinespace[0.3em]
\multicolumn{3}{l}{\textbf{Inefficient Allocation of Space
    (Section \ref{inefficient_allocation})}}\\
\hspace{1em}background $\ge\frac{1}{3}$ & 82.5 & $\pm$0.5 & 23\% & $\pm$0.1\%\\
\addlinespace[0.3em]
\multicolumn{3}{l}{\textbf{Completeness
    (Section \ref{missing_elements})}}\\
\hspace{1em}yes & 105~~ & $\pm$2 & 29\% & $\pm$0.6\%\\
\em{\hspace{1em}no} & \em{257~~} & \em{$\pm$2} & \em{71\%} & \em{$\pm$0.6\%}\\
\addlinespace[0.3em]
\multicolumn{3}{l}{\textbf{Informativeness Gap
    (Section \ref{igaps})}}\\
\hspace{1em}no & 162.5 & $\pm$40.5 & 45\% & $\pm$11\%\\
\em{\hspace{1em}yes} & \em{199.5} & \em{$\pm$40.5} & \em{55\%} & \em{$\pm$11\%}\\
\addlinespace[0.3em]
\multicolumn{3}{l}{\textbf{Announcements
    (Section \ref{announcements})}}\\
\hspace{1em}no & 274~~ & $\pm$2 & 76\% & $\pm$0.6\%\\
\em{\hspace{1em}yes} & \em{88~~} & \em{$\pm$2} & \em{24\%} & \em{$\pm$0.6\%}\\
\addlinespace[0.3em]
\multicolumn{3}{l}{\textbf{Understandability Gap
    (Section \ref{ugaps})}}\\
\hspace{1em}no & 305~~ & $\pm$38 & 84\% & $\pm$10\%\\
\em{\hspace{1em}yes} & \em{57~~} & \em{$\pm$38} & \em{16\%} & \em{$\pm$10\%}\\
\addlinespace[0.3em]
\multicolumn{3}{l}{\textbf{Ambiguous Formulations
    (Section \ref{ambiguity})}}\\
\hspace{1em}no & 323~~ &  & 89\% & \\
\em{\hspace{1em}yes} & \em{39~~} & \em{} & \em{11\%} & \em{}\\
\addlinespace[0.3em]
\multicolumn{3}{l}{\textbf{Properness
    (Section \ref{hownotto_summary})}}\\
\hspace{1em}yes & 13~~ & $\pm$6 & 3.6\% & $\pm$1.7\%\\
\em{\hspace{1em}no} & \em{349~~} & \em{$\pm$6} & \em{96\%} & \em{$\pm$1.7\%}\\
\bottomrule
\end{tabular}
\end{table}

\end{knitrout}

\subsection{How To: Proper Abstracts}\label{howto_proper}

Indeed, the proper abstracts are so few, we can list them here completely.
\begin{itemize} % #nonknitr
\item Empirical works with structured abstract:
  \Art{AmnPoe22}, \Art{LavMor22}, \Art{TanFeiAvg22}, \Art{YuKeuXia22}.
\item Empirical works with non-structured abstract:
  \Art{AbdBadCos22}, \Art{CinCooPas22}, \Art{HeMenChe22}, \Art{LinWilHal22}, \Art{OliAssGar22}, 
\Art{UddAlaSer22}, \Art{WanZhaZha22}.
\item Design works with structured abstract:
  \Art{BaiJiaCap22}, \Art{WuSheChe22}.
\item Design works with non-structured abstract:
  \Art{BarDuDav22}, \Art{CorRweFra22}, \Art{GalEwaJun22}, \Art{GerMarLat22}, \Art{HanMeh22}, \Art{YeGuMar22}.
\end{itemize}

\subsection{How To: Information-Rich Formulations}

When avoiding the uninformative formulations of Section~\ref{uninformative},
some authors manage to pack large amounts of relevant information into a sentence.
This most often occurs for \Cb{method} or \Cb{result} information.

Consider this \Cb{method} sentence:
\Quote{Therefore, through a case study of 9 open source software projects across 30 versions, 
  we study the relative effectiveness of SNA metrics when compared to code metrics across 
  3 commonly used SDP contexts (Within-project, Cross-version and Cross-project) 
  and scenarios (Defect-count, Defect-classification (classifying if a module is defective) and Effort-aware (ranking
the defective modules w.r.t to the involved effort)).} (\Art{GonRajHas22})

This sentence is complex, but pays back for the reading effort by being \emph{enormously} informative.
 
Or consider this well-designed \Cb{method}+\Cb{result} sentence:
\Quote{We evaluate the three versions on a set of 299 data constraints from 15 real-world Java
  systems, and find that they improve method-level link recovery
  by 30\%, 70\%, and 163\%, in terms of true positives within the first
  10 results, compared to their text-retrieval-based baseline.} (\Art{FloPerWei22})

Easily readable, related information is kept together, and the meaning of results is explained precisely.
We wished more authors would write like this!

\subsection{How To: Structured Abstracts are more orderly}\label{structuredgood}

This study is mostly exploratory.
Our only clear expectation at the start was that structured abstracts would tend to be
better (whatever that was going to mean) than unstructured ones.
Therefore, we perform two statistical hypothesis tests that compare
structured to unstructured abstracts.

We find that structured abstracts are significantly more often \emph{complete}
($\chi^2=49.9, p<0.001$).
For \emph{properness}, the absolute numbers are so low that statistical significance
is not achieved
($\chi^2=0.5, p=0.48$)
despite the trend visible in the
relative difference of the two green bars in the right half of
Figure~\ref{nonzerofractionbar_xletgroups_totalqualityfractions}.
Nevertheless, overall the expectation appears to be correct.

%========================================================================
\section{Limitations and Threats to Validity}

\subsection{Interpretivist Perspective}

In terms of Tracy's quality criteria for qualitative research \cite{Tracy10},
our study has nice properties, because our readers inhabit
the domain of abstract reading and abstract writing themselves.

The topic's worthiness is obvious, the amount of data is large,
as is the number of constructs used.
Challenges are discussed below.
Hopefully, our constructs and findings resonate with you;
they certainly resonate with us.
Should you find our description insufficiently thick,
you can easily look up lots of additional examples in our raw data
described in Section~\ref{dataavailability}.

\subsection{Positivist Perspective}

Internal validity is the degree to which the stated method
was followed correctly.
We do not expect much problem in this regard.
The most difficult-to-avoid problem in our study is lapses of concentration
during coding which result in coding mistakes.
However, the laborious coding procedure
described in Section~\ref{meth_coding}
makes it very unlikely for such mistakes to slip through.
Most other steps were automated, so mistakes would be systematic
and not likely to escape our attention.

Construct validity is the degree to which the design
of the study is adequate for the phenomenon to be understood.
Here, our study has obvious limitations:
It ideally ought to measure the informativeness and understandability of
abstracts. However, both of these are reader-dependent, so
measuring them would involve a reading study with many readers.
This would face huge problems in getting a representative set of readers,
could never scale to the hundreds of abstracts we look at here,
and, whichever operationalization it chose,
it would be imperfect and controversial.
We therefore decided to analyze properties of abstracts
that are \emph{arguably} problematic, although we cannot say just how
problematic in each case, resulting in count statistics only.

External validity is the degree to which the results generalize
to other sets of abstracts:
Here, we expect that our results generalize well to
neighboring (past and future) years in the same venues
we studied, perhaps a bit less for ICSE because of its
varying location and hence more varying authors.
Whether it also generalizes to other venues
we cannot know, but we would be surprised if venues of lower
scientific reputation had articles with much better abstracts.

%========================================================================
\section{Conclusions}

\subsection{Too Few Abstracts Have Good Quality}

Only 29\% of the investigated software engineering research article abstracts are 
\emph{complete} according to the IMRAD structure long established for abstracts in science
(Figure~\ref{nonzerofractionbar_xletgroups_totalqualityfractions}).
In particular, more than half of all abstracts
(62\%) never formulate a conclusion in the sense of
a generalizing take-home message (Figure~\ref{zerofractionbar_xletgroups_topicmissingfractions}).
Only 4\% of the abstracts are \emph{complete}
and also fulfill modest additional criteria of 
informativeness and understandability 
(Figures~\ref{boxplots_fkscore} and \ref{nonzerofractionbar_xletgroups_missinginfofractions}, 
Section~\ref{ambiguity}).
We define and quantify further quality issues not included in the above number 
(Figures~\ref{box_xletgroups_topicfractions}, \ref{box_xletgroups_conclusionfractions}, 
\ref{nonzerofractionbar_xletgroups_missinginfofractions}, and non-quantitative
Section~\ref{confusing}).
This low abstracts quality exists despite the fact that we analyzed only venues
supposed to have high quality.

This is deplorable, because for the vast majority of readers,
the abstract is the first part they read \cite{ShiGalMil24} --- and often the only part.
Futhermore, scientific indexes and specialized search engines such as \emph{Scopus AI}\footnote{%
  \url{https://www.elsevier.com/products/scopus/scopus-ai}
} rely heavily on abstracts.
With the above level of abstracts quality, readers will get far less
well informed than they could have been and the spread of knowledge
will be slowed down accordingly, needlessly wasting public money.

We conclude that the software engineering research community should pay
more attention to abstract-writing.
Presumably, introducing a structured abstract format 
and accompanying writing instructions that suit 
engineering research would help.

\subsection{How To: Guidelines for Well-Written Abstracts}\label{howto-guidelines}

\subsubsection{For Authors}\label{guidelinesauthors}

The steps cross-reference the article sections that supply 
detail or evidence.

\begin{enumerate}
  \item Write a structured abstract, not a free-flowing one (see Section~\ref{structuredgood}) 
    and follow the archetype (Sections~\ref{archetype}).
    If the venue requires structured abstracts in an unsuitable format, protest.
  \item Take care to avoid announcements (see Section~\ref{announcements}),
    understandability gaps (see Section~\ref{ugaps}),
    and sentences with an unclear role (see Section~\ref{ambiguity}).
    Provide helpful detail, perhaps simplified, if it requires only little space
    (see Section~\ref{igaps}).
    Keep your sentences short.
  \item Write a short \Sah{Background} section that provides just enough context
    and motivation to understand the subsequent \Sah{Objective} 
    (see Section~\ref{inefficient_allocation}).
  \item Decide on your main contribution. 
    Is your article a design article or an empirical one?
    Write an \Sah{Objective} that expresses the type and your specific goal succinctly
    (see Section~\ref{confusing}).
    If your background information contains a corresponding \Sah{Gap} statement,
    clearly phrase it as such.
  \item If your work is a design article, 
    write a \Sah{Design} section.
    Cover all key ideas (typically two to four), avoid non-key information.
  \item If you have multiple near-independent empirical sub-studies and
    your work is a design article,
    use two or three combined \Sah{Method and Results} sections.
    Each of these will typically be a single sentence of the form
    ``We do-this-and-that and find this-and-that.''
    (see Section~\ref{confusing}).
    Otherwise write separate \Sah{Methods} and \Sah{Results} sections
    as follows.
  \item Write a \Sah{Methods} section that explains what you have done
    for your empirical study. 
    Be as specific as you can do concisely.
    In particular, mention the amount of data used (see Section~\ref{igaps}).
  \item Write a \Sah{Results} section that explains the main outcomes
    of your study.
    If you have many results, report the one or two most important ones.
    If you have many results of equal importance, report the 
    one or two most interesting ones or just provide examples.
    Be specific and beware of announcements (see Section~\ref{announcements}).
  \item Write a \Sah{Conclusion} section that generalizes from your results.
    What should the reader take home?
    What do we now know that we did not know before?
    The broader the conclusion, the higher the relevance of your work,
    but the lower its credibility. 
    Find a formulation with good relevance and good-enough credibility.
    Do not be a coward (see Section~\ref{missing_elements}):
    Dare to later be proven wrong for a few of your works.
  \item An outlook on future possibilities \emph{can} be part of your \Sah{Conclusion},
    but is usually better left to the body of your article.
\end{enumerate}

\subsubsection{For Venues}

Structured abstracts are not currently used widely in software engineering, presumably
because their usual form does not suit design articles. 

In a suitably extended form, however, 
structured abstracts promise better abstracts quality than a free-style format
because our results strongly suggest that a structured format is helpful for 
completeness. Previous results show that it is also helpful for understandability.
Therefore, venues should require structured abstracts in such a new,
engineering-ready format as described by 
Table~\ref{tab:codes} and Sections~\ref{archetype} and \ref{guidelinesauthors}:
allowing \Sah{Gap} statements, 
allowing \Sah{Design} sections, and  
allowing multiple \Sah{Methods and Results} sections.

For your call for papers, feel free to copy the above text,
point your readers to the online version,\footnote{\url{https://github.com/serqco/qabstracts/blob/main/serqco-abstracts-structure.md}}
perhaps point authors to this article for the underlying evidence.

%========================================================================

\subsection*{Acknowledgments}
\noindent We thank Gesine Milde for cleansing the automatically extracted abstract texts.

\bibliographystyle{IEEEtran}
\bibliography{special.bib}

% Generated by IEEEtran.bst, version: 1.14 (2015/08/26)
\begin{thebibliography}{10}
\providecommand{\url}[1]{#1}
\csname url@samestyle\endcsname
\providecommand{\newblock}{\relax}
\providecommand{\bibinfo}[2]{#2}
\providecommand{\BIBentrySTDinterwordspacing}{\spaceskip=0pt\relax}
\providecommand{\BIBentryALTinterwordstretchfactor}{4}
\providecommand{\BIBentryALTinterwordspacing}{\spaceskip=\fontdimen2\font plus
\BIBentryALTinterwordstretchfactor\fontdimen3\font minus
  \fontdimen4\font\relax}
\providecommand{\BIBforeignlanguage}[2]{{%
\expandafter\ifx\csname l@#1\endcsname\relax
\typeout{** WARNING: IEEEtran.bst: No hyphenation pattern has been}%
\typeout{** loaded for the language `#1'. Using the pattern for}%
\typeout{** the default language instead.}%
\else
\language=\csname l@#1\endcsname
\fi
#2}}
\providecommand{\BIBdecl}{\relax}
\BIBdecl

\bibitem{Lang22}
T.~A. Lang, ``Scientific abstracts: Texts, contexts, and subtexts,''
  \emph{European Science Editing}, vol.~48, 2022.

\bibitem{ShiGalMil24}
F.~Shiely, K.~Gallagher, and S.~R. Millar, ``How, and why, science and health
  researchers read scientific ({IMRAD}) papers,'' \emph{PLOS One}, vol.~19,
  no.~1, 2024.

\bibitem{StrCor90}
A.~Strauss and J.~Corbin, \emph{Basics of qualitative research}.\hskip 1em plus
  0.5em minus 0.4em\relax Sage Publications, 1990.

\bibitem{Krippendorff04}
K.~Krippendorff, \emph{Content analysis: An introduction to its methodology},
  2nd~ed.\hskip 1em plus 0.5em minus 0.4em\relax Sage Publications, 2004.

\bibitem{Swales90}
J.~M. Swales, \emph{Genre analysis: English in academic and research
  settings}.\hskip 1em plus 0.5em minus 0.4em\relax Cambridge University Press,
  1990.

\bibitem{DosSantos96}
M.~B. Dos~Santos, ``The textual organization of research paper abstracts in
  applied linguistics,'' \emph{Text \& Talk}, vol.~16, no.~4, pp. 481--500,
  1996.

\bibitem{CroOpp06}
C.~Cross and C.~Oppenheim, ``A genre analysis of scientific abstracts,''
  \emph{Journal of Documentation}, vol.~62, no.~4, pp. 428--446, 2006.

\bibitem{SolPer04}
L.~B. Sollaci and M.~G. Pereira, ``The introduction, methods, results, and
  discussion ({IMRAD}) structure: a fifty-year survey,'' \emph{Journal of the
  Medical Library Association}, vol.~92, no.~3, p. 364, 2004.

\bibitem{DupKhoLeb03}
A.~Dupuy, K.~Khosrotehrani, C.~Lebb{\'e}, M.~Rybojad, and P.~Morel, ``Quality
  of abstracts in 3 clinical dermatology journals,'' \emph{Archives of
  Dermatology}, vol. 139, no.~5, pp. 589--593, 2003.

\bibitem{ShaHar06}
S.~Sharma and J.~E. Harrison, ``Structured abstracts: do they improve the
  quality of information in abstracts?'' \emph{American Journal of Orthodontics
  and Dentofacial Orthopedics}, vol. 130, no.~4, pp. 523--530, 2006.

\bibitem{RosGreSto05}
A.~B. Rosen, D.~Greenberg, P.~W. Stone, N.~V. Olchanski, and P.~J. Neumann,
  ``Quality of abstracts of papers reporting original cost-effectiveness
  analyses,'' \emph{Medical Decision Making}, vol.~25, no.~4, pp. 424--428,
  2005.

\bibitem{FagLiuHud14}
C.~M. Faggion, Jr, J.~Liu, F.~Huda, and M.~Atieh, ``Assessment of the quality
  of reporting in abstracts of systematic reviews with meta-analyses in
  periodontology and implant dentistry,'' \emph{Journal of Periodontal
  Research}, vol.~49, no.~2, pp. 137--142, 2014.

\bibitem{ArtZaaChe20}
W.~Arthur, Z.~Zaaza, J.~X. Checketts, A.~L. Johnson, K.~Middlemist, C.~Basener,
  S.~Jellison, C.~Wayant, and M.~Vassar, ``Analyzing spin in abstracts of
  orthopaedic randomized controlled trials with statistically insignificant
  primary endpoints,'' \emph{Arthroscopy: The Journal of Arthroscopic \&
  Related Surgery}, vol.~36, no.~5, pp. 1443--1450, 2020.

\bibitem{ReyRidBro20}
V.~Reynolds-Vaughn, J.~Riddle, J.~Brown, M.~Schiesel, C.~Wayant, and M.~Vassar,
  ``Evaluation of spin in the abstracts of emergency medicine randomized
  controlled trials,'' \emph{Annals of Emergency Medicine}, vol.~75, no.~3, pp.
  423--431, 2020.

\bibitem{JelRobBow20}
S.~Jellison, W.~Roberts, A.~Bowers, T.~Combs, J.~Beaman, C.~Wayant, and
  M.~Vassar, ``Evaluation of spin in abstracts of papers in psychiatry and
  psychology journals,'' \emph{BMJ Evidence-Based Medicine}, vol.~25, no.~5,
  pp. 178--181, 2020.

\bibitem{BouDutRav10}
I.~Boutron, S.~Dutton, P.~Ravaud, and D.~G. Altman, ``Reporting and
  interpretation of randomized controlled trials with statistically
  nonsignificant results for primary outcomes,'' \emph{JAMA}, vol. 303, no.~20,
  pp. 2058--2064, 2010.

\bibitem{MohHopSch12}
D.~Moher, S.~Hopewell, K.~F. Schulz, V.~Montori, P.~C. G{\o}tzsche, P.~J.
  Devereaux, D.~Elbourne, M.~Egger, and D.~G. Altman, ``{CONSORT} 2010
  explanation and elaboration: updated guidelines for reporting parallel group
  randomised trials,'' \emph{International Journal of Surgery}, vol.~10, no.~1,
  pp. 28--55, 2012.

\bibitem{HarSydBlu96}
J.~Hartley, M.~Sydes, and A.~Blurton, ``Obtaining information accurately and
  quickly: are structured abstracts more efficient?'' \emph{Journal of
  Information Science}, vol.~22, no.~5, pp. 349--356, 1996.

\bibitem{KitDybJor04}
B.~A. Kitchenham, T.~Dybå, and M.~Jørgensen, ``Evidence-based software
  engineering,'' in \emph{Proc. 26th Int'l. Conf. on Software
  Engineering}.\hskip 1em plus 0.5em minus 0.4em\relax IEEE, 2004, pp.
  273--281.

\bibitem{KitBreOwe08}
B.~A. Kitchenham, O.~P. Brereton, S.~Owen, J.~Butcher, and C.~Jefferies,
  ``Length and readability of structured software engineering abstracts,''
  \emph{IET Software}, vol.~2, no.~1, pp. 37--45, 2008.

\bibitem{BudKitCha08}
D.~Budgen, B.~A. Kitchenham, S.~M. Charters, M.~Turner, P.~Brereton, and S.~G.
  Linkman, ``Presenting software engineering results using structured
  abstracts: a randomised experiment,'' \emph{Empirical Software Engineering},
  vol.~13, pp. 435--468, 2008.

\bibitem{PreMonFra25}
L.~Prechelt, L.~Montgomery, J.~Frattini, and F.~Zieris, ``How (not) to write a
  software engineering abstract (data package),'' zenodo.org,
  DOI:10.5281/zenodo.15736253, \url{https://doi.org/10.5281/zenodo.15736252},
  2025.

\bibitem{KinFisRog75}
\BIBentryALTinterwordspacing
J.~P. Kincaid, R.~P. Fishburne, Jr, R.~L. Rogers, and B.~S. Chissom,
  ``Derivation of new readability formulas ({Automated Readability Index}, {Fog
  Count} and {Flesch Reading Ease Formula}) for navy enlisted personnel,''
  United States. Naval Education and Training Support Command. Chief of Naval
  Technical Training, Research Branch Report 8-75, 1975. [Online]. Available:
  \url{https://stars.library.ucf.edu/istlibrary/56/}
\BIBentrySTDinterwordspacing

\bibitem{Tracy10}
S.~J. Tracy, ``Qualitative quality: Eight “big-tent” criteria for excellent
  qualitative research,'' \emph{Qualitative Inquiry}, vol.~16, no.~10, pp.
  837--851, 2010.

\end{thebibliography}

\begin{IEEEbiography}[{\includegraphics[width=1in,height=1.25in,clip,keepaspectratio]{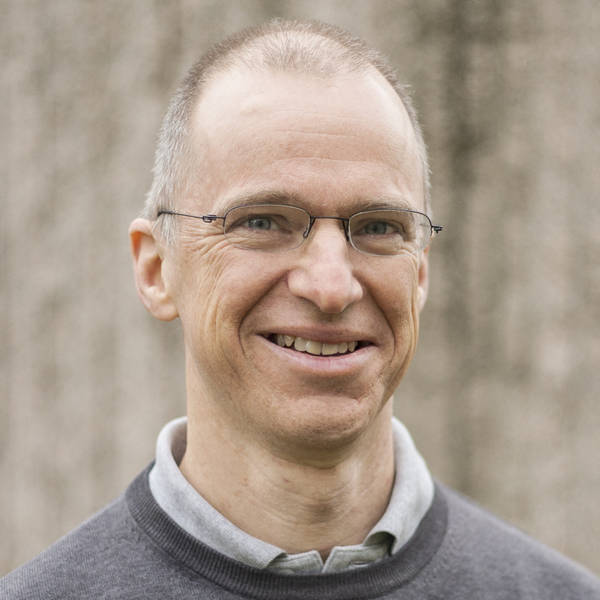}}]{Lutz Prechelt}
received a PhD from the University of Karlsruhe
for work that combined machine learning and compiler construction
for parallel machines.
He then moved to empirical software engineering and performed
a number of controlled experiments before spending three
years in industry as an engineering manager and CTO.
He is now full professor for software engineering at
Freie Universität Berlin and executive director of the
Institute for Informatics there.
His research interests concern the human
factor in the software development process, asking mostly
exploratory research questions and addressing them with
qualitative methods.
Additional research interests concern research methods
and the health of the research system.
He is the founder of the 
Software Engineering Research Quality Coalition (SERQco)
and the inventor of Review Quality Collector (RQC).
Contact him at prechelt@inf.fu-berlin.de.
\end{IEEEbiography}

\begin{IEEEbiography}[{\includegraphics[width=1in,height=1.25in,clip,keepaspectratio]{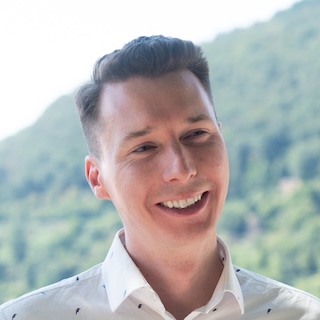}}]{Lloyd Montgomery}
received his PhD from the University of Hamburg, Germany, working under Prof. Dr. Walid Maalej.
Currently, his primary research area is the quality of issue tracking systems, with other interests such as empirical software engineering, non-technical factors in software engineering, and science communication.
He won the RE'17 best paper award for his machine learning design science research with IBM customer support.
Lloyd's academic service record includes serving as the IST Publicity Chair, RE'23 Publicity Chair, NLP4RE'22 Workshop Co-Chair, and RE'21 Artefact Co-Chair.
He is on the program committees of REFSQ, NLP4RE, and the RE@Next! track at RE, and regularly reviews for the journals REJ, IST, and JSS, in addition to reviews for SQ Journal and TOSEM.
Lloyd also has a particular passion for artefact tracks, having served as a PC member for seven artefact tracks at RE, ICSE, and FSE.
Contact him at lloyd.montgomery@uni-hamburg.de.
\end{IEEEbiography}  

\begin{IEEEbiography}[{\includegraphics[width=1in,height=1.25in,clip,keepaspectratio]{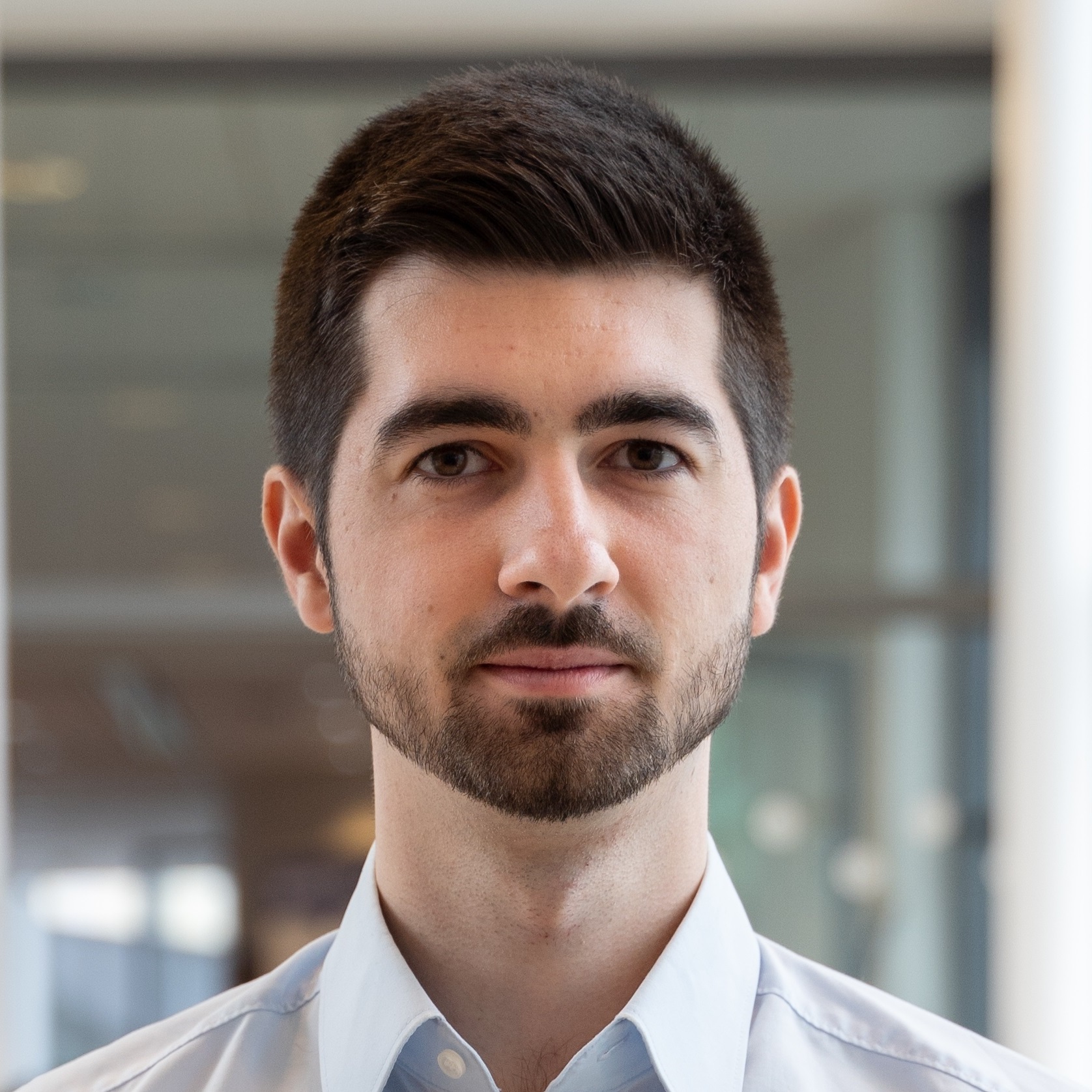}}]{Julian Frattini}
(julian.frattini@chalmers.se) obtained his Ph.D. degree from Blekinge Institute of Technology (BTH), Sweden, and is currently a postdoctoral researcher at Chalmers University of Technology and University of Gothenburg, Sweden.
He contributes research to requirements engineering with his work on requirements artifact quality, as well as to the field of software engineering research methodology with a particular interest in statistical causal inference and Bayesian data analysis.
Julian served on the organizing committees of REFSQ, RE, AIRE, and CrowdRE, and is the publicity co-chair of the IST journal.
He is on the program committees of RE, REFSQ, EASE, ESEM, QUATIC, and NLP4RE, and reviews for the journals TSE, TOSEM, EMSE, REJ, IST, and JSS.
\end{IEEEbiography}

\begin{IEEEbiography}[{\includegraphics[width=1in,height=1.25in,clip,keepaspectratio]{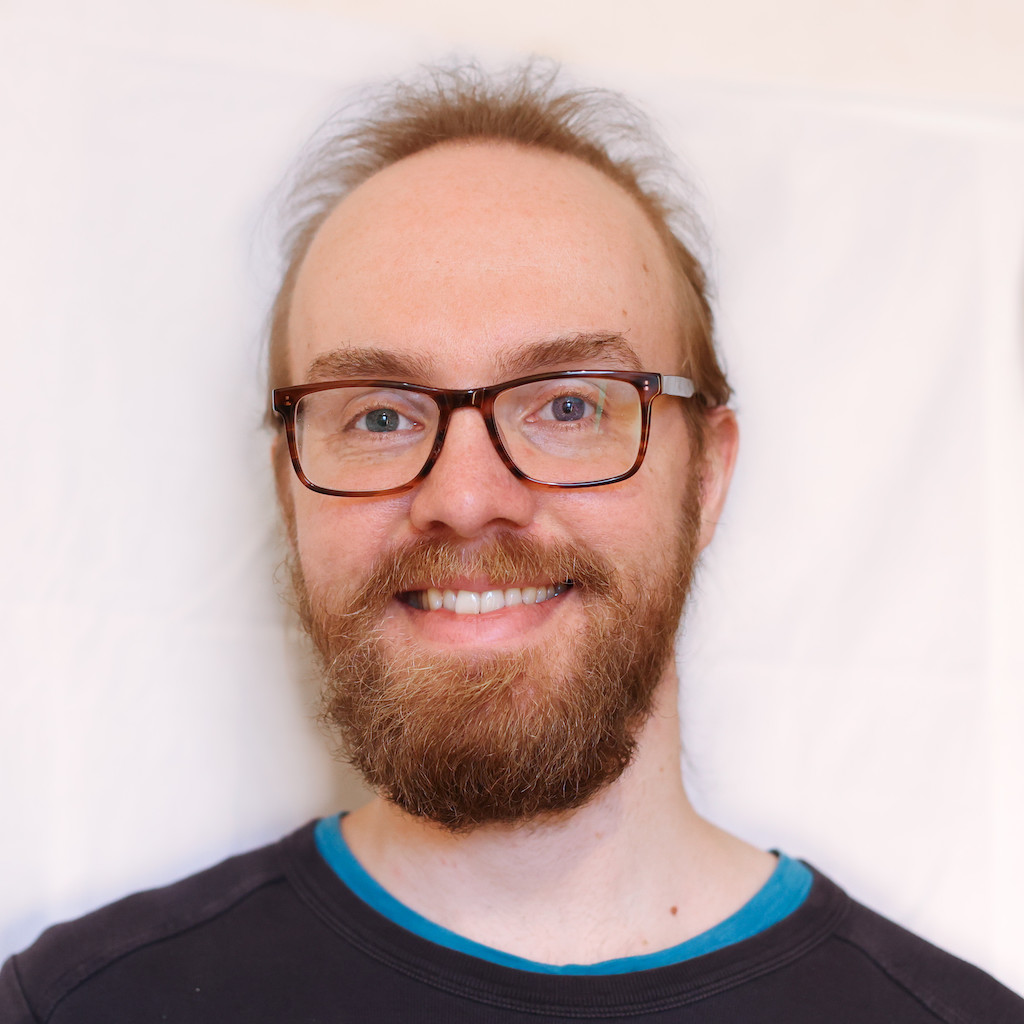}}]{Franz Zieris}
received a PhD from Freie Universität Berlin for qualitative research on pair programming in industy.
He then spent 2.5 years in industry as software architect and business analyst.
He is now an Associate Senior Lecturer at the Department of Software Engineering at BTH,
where his research focuses on continuous software engineering and distributed software engineering.
Franz is on the program committes of the Internal Conference on Agile Software Development (XP)
and International Conference on Cooperative and Human Aspects (CHASE),
and regularly reviews for Empirical Software Engineering (EMSE),
the Journal of Systems and Software (JSS),
and Information and Software Technology (IST).
\end{IEEEbiography}

\ifwithappendix{\input{appendix}}

\end{document}